\RequirePackage{ifpdf}

\documentclass[a4paper,11pt]{article}
\pdfoutput=1 % if your are submitting a pdflatex (i.e. if you have
\usepackage{jheppub} % for details on the use of the package, please
\usepackage[T1]{fontenc} % if needed

\usepackage{ifpdf} % to get it to work with pdflatex and not break for everything else

\usepackage{flexisym}
\usepackage{breqn}
\usepackage{comment}

\author{Costas G. Papadopoulos}

\affiliation{Institute of Nuclear and Particle Physics, NCSR `Demokritos', Agia Paraskevi, 15310, Greece}

\emailAdd{costas.papadopoulos@cern.ch}


\title{Simplified differential equations approach for Master Integrals}

\abstract{A simplified differential equations approach for Master Integrals is presented. It allows to express 
them, straightforwardly,  in terms of Goncharov Polylogarithms. As a proof-of-concept of the proposed method, results at one and two loops are
presented, including the massless one-loop pentagon with up to one off-shell leg  at order epsilon.}

\ifpdf
\fi

\newcommand{\eqn}[1]{Eq.~\ref{#1}}
\newcommand{\be}{\begin{equation}}
\newcommand{\ee}{\end{equation}}
\newcommand{\ba}{\begin{array}}
\newcommand{\ea}{\end{array}}
\newcommand{\bea}{\begin{eqnarray}}
\newcommand{\eea}{\end{eqnarray}}

\newcommand{\bqa}{\begin{eqnarray}}
\newcommand{\eqa}{\end{eqnarray}}

\begin{document}
\unitlength1cm
\maketitle
\allowdisplaybreaks

\section{Introduction}

With the advance of LHC program, precise theoretical predictions for scattering processes become indispensable. The last years 
next-to-leading order (NLO) calculations~\cite{vanHameren:2009dr,Berger:2009zg,Bern:2013gka,Bredenstein:2009aj,Bevilacqua:2009zn,Bevilacqua:2010ve} 
have been automatized  and become a very valuable tool for the physics analysis of the
experimental data. Next-to-next-to-leading order (NNLO) calculations are also of paramount importance for efficiently exploring the 
available and forthcoming data. Many of the NNLO calculations are already heavily used, especially those for Higgs~\cite{Anastasiou:2002yz} and Vector Boson production~\cite{Anastasiou:2003ds}.

At the NLO level, a general decomposition of the one-loop amplitude in terms of scalar one-loop
integrals, the Master Integrals (MI) at the one-loop level, was known 
long ago~\cite{Melrose:1965kb,'tHooft:1978xw,Passarino:1978jh,vanNeerven:1983vr,Bern:1992em}. 
Combined with the development of reduction at the 
integrand level techniques~\cite{Ossola:2006us,Ossola:2008xq,Giele:2008ve,Berger:2008sj,Mastrolia:2012bu}, resulted to the full automation of the NLO calculations: many software packages are nowadays available that may 
accomplish the task of computing NLO corrections to arbitrary processes~\cite{Bevilacqua:2011xh,Ossola:2007ax,Giele:2008bc,Hirschi:2011pa,Cullen:2011ac,Bern:2013pya,Badger:2010nx,Cascioli:2011va,Actis:2013dfa}. 

Beyond one loop, calculations are notoriously difficult, especially if an exclusive description of the
scattering processes is needed. During the last couple of years important progress has been achieved at the NNLO frontier. First
attempts towards a generalization of unitarity~\cite{Gluza:2010ws,Kosower:2011ty}  and reduction at the integrand level methods~\cite{Mastrolia:2011pr,Badger:2012dp,Zhang:2012ce,Badger:2013gxa} at
two loops for practical calculations have been initiated offering new ground for related developments.
A first attempt for the construction of integrand basis in $d = 4$ has already been presented in \cite{Feng:2012bm}.
Algebraic geometry offered valuable tools (for instance the concept of Groebner basis in relation with
multivariate polynomial division) to understand the reduction at a deeper level. It is by now clear that a
general solution to the problem of integrand level reduction is within reach ~\cite{Mastrolia:2013kca,Papadopoulos:2013hra}. The new insights
in the analytic structure of MI involved in the calculation of two-loop amplitudes offered by the
algebra of symbols of (certain) transcendental (multiple polylogarithms) functions~\cite{Goncharov:1998kja,Goncharov:2001,Goncharov:2010jf,Duhr:2011zq,Duhr:2012fh}, resulted
to several advances towards the completion of the integral basis at two loops~\cite{Chavez:2012kn,Gehrmann:2013cxs}. For the double-real
contributions a new algorithm has been proposed~\cite{Czakon:2010td}, that seems to provide a general solution. 
Moreover, progress in the antenna subtraction method
has been presented in \cite{GehrmannDeRidder:2012ja}. Developments relevant to real-virtual contributions~\cite{Boughezal:2011jf} have been appeared~\cite{Bierenbaum:2011gg}
completing the known solution for massless amplitudes~\cite{Bern:1998sc,Bern:1999ry,Catani:2000pi}. 
First NNLO QCD predictions for $2 \to 2$
scattering processes at hadron colliders have appeared recently, namely for (all-gluon)
di-jet~\cite{Ridder:2013mf}, $g g\to H+$jet~\cite{Boughezal:2013uia} and the complete $t\bar{t}$~\cite{Czakon:2013goa} production. 
The successful completion of these calculations has been based on
novel advanced algorithms to tackle with NNLO real corrections with two unresolved patrons in the
final state. Nevertheless a general solution for NNLO calculations in the same lines as for the NLO
ones, is not yet available, although all evidences point to the fact that this accomplishment is within
reach over the next few years, resulting to the next breakthrough in applied quantum field theory
calculations.

One of the main problems in extending the successful NLO approach at the NNLO level remains the
evaluation of the full set of MI involved.  In the last fifteen years, the calculation of virtual
corrections has been revolutionized with the 
 advent of automated reduction techniques to MI~\cite{Chetyrkin:1981qh,Tkachov:1981wb,Laporta:2001dd,Smirnov:2004ym}
 and the development of  
systematic solutions of differential equations~\cite{Kotikov:1990kg,Bern:1992em,Bern:1993kr,Gehrmann:1999as} satisfied by MI or the evaluation of their Mellin-Barnes
representations~\cite{Smirnov:1999gc,Tausk:1999vh}.
%with the use of the algebra properties of  harmonic polylogarithms. 

The differential equations approach (DE) has proven to be very powerful in a large number
of computations, including two-loop four-point functions with massless and massive
internal propagators. Within this framework, DE for the MI are derived, in terms of kinematical invariants.
The method relies heavily on the use of integration-by-parts identities (IBPI) that allows to reduce all integrals
involved to a relatively small subset of MI.  
The MI are then evaluated by solving these DE, matched to appropriate boundary conditions.
One of the important issues is to find the proper parametrization of the kinematics involved, that simplifies 
the form of the DE and allows the expression of their solution in terms of known analytic functions, that are then
easily computable. In that respect the use of iterated integrals, more specifically of Goncharov Polylogarithms (GPs)~\cite{Goncharov:1998kja,Goncharov:2001,Goncharov:2010jf,Ablinger:2013cf}, 
is of paramount importance, at least for MI with vanishing internal masses, a very important class of virtual 
corrections related to NNLO QCD calculations.

In this paper we present the first steps towards the establishment of DE that are simple enough and their results are
straightforwardly expressible in terms of GPs. In order to illustrate the idea we present in detail, in section \ref{section:DE}, how the method
works at one and two loops.
At one loop we complete the calculation of five-point function~\cite{Bern:1993kr} at order $\varepsilon$, which, to the best of our knowledge, is a new result~\cite{DelDuca:2009ac}.
At two loops we derive, as a proof-of-concept of our method, results for three-point MI with three off-shell legs and for certain four-point MI with two off-shell legs.
 Finally, in section \ref{section:Di}, we summarize our findings and discuss the open issues.

\section{Differential equations}
\label{section:DE}

Originally differential equations have been derived for MI by differentiating with respect to kinematical invariants and then using IPBI in order to express their derivatives 
in terms of other MI of equal or less complexity. 
The proposal in this paper is to formulate differential equations with respect to a parameter, that are simple enough and straightforwardly solvable in terms of GPs.
We will illustrate the idea by examples at one and two loops.

\subsection{One-loop results}

Let us start by defining the one-loop MI.
The general $n-$point integral, with vanishing internal masses, is defined by
\be
\int {\frac{{d^d k}}{{i\pi ^{d/2} }}\frac{1}{{D_0 D_1  \ldots D_{n - 1} }}} 
\ee
with $D_i  =  - \left( {k + p_0  +  \ldots  + p_i } \right)^2$ and take for convenience $p_0=0$.
It can be considered as a function of the external momenta $p_i$. It belongs to the topology defined by
\[
G_{a_1  \ldots a_n }  = \int {\frac{{d^d k}}{{i\pi ^{d/2} }}\frac{1}{{D_0^{a_1 } D_1^{a_2 }  \ldots D_{n - 1}^{a_n } }}} 
\]
namely $G_{1...1}$.
We now introduce a simple parametrization as follows:
\be
G_{11...1} (x) = \int {\frac{{d^d k}}{{i\pi ^{d/2} }}\frac{1}{{\left( { - k^2 } \right)\left( { - \left( {k + x\,p_1 } \right)^2 } \right)\left( { - \left( {k + p_1  + p_2 } \right)^2 } \right) \ldots \left( { - \left( {k + p_1  + p_2  +  \ldots  + p_n } \right)^2 } \right)}}} 
\ee
Now the integral becomes a function of $x$, which allows to define a differential equation with respect to $x$, schematically given by
\be
\frac{\partial }{{\partial x}}G_{11 \ldots 1} \left( x \right) =  - \frac{1}{x}G_{11 \ldots 1} \left( x \right) + x p_1 ^2 G_{12...1}  + \frac{1}{x}G_{02...1} 
\ee
Using IBPI~\cite{Smirnov:2008iw} the r.h.s. can be expressed as a sum of MI of lower complexity multiplied by rational functions of $x$. 
Iterating the procedure we would like to express any MI defined above as a sum of GPs.

\begin{figure}
\begin{center}
\includegraphics[width=8cm]{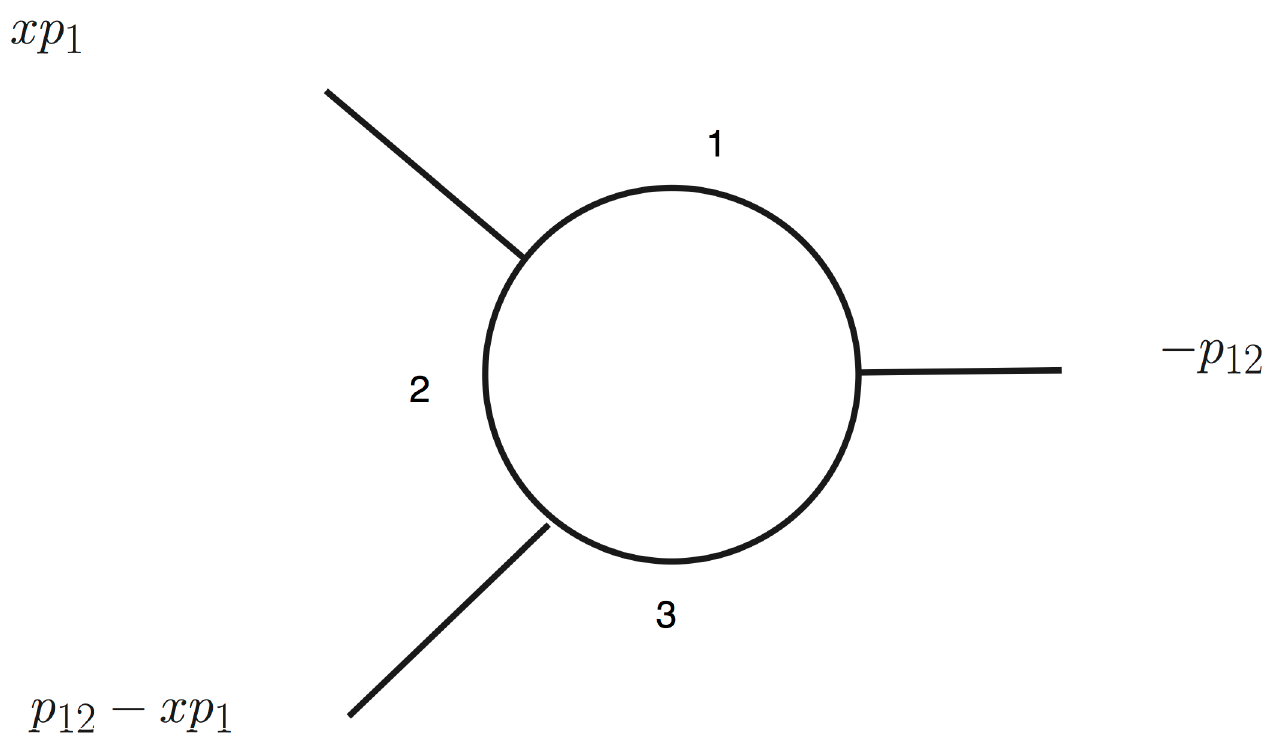}
%\[ \epsfig{figure=one3m.eps,width=8cm} \]
\caption{The one-loop graph with three off-shell legs.The label $1,2,3$ refer to the denominators $-(k_1)^2,-(k_1+x p_1)^2,-(k_1+p_1+p_2)^2$, \eqn{1l3m}.}
\label{fig:1l3p}
\end{center}
\end{figure}

To be more specific we consider first the $3-$point integral with off-shell legs given in Fig.~\ref{fig:1l3p}
 \be
T(q_1^2 ,q_2^2 ,q_3^2 ) = \int {\frac{{d^d k}}{{i\pi ^{d/2} }}\frac{1}{{\left( { - k^2 } \right)\left( { - \left( {k + q_1 } \right)^2 } \right)\left( { - \left( {k + q_1  + q_2 } \right)^2 } \right)}}} 
\ee
which we  parametrize as
\be
\begin{array}{ll}
 G_{111} (x) &= \int {\frac{{d^d k}}{{i\pi ^{d/2} }}\frac{1}{{\left( { - k^2 } \right)\left( { - \left( {k + x\,p_1 } \right)^2 } \right)\left( { - \left( {k + p_1  + p_2 } \right)^2 } \right)}}}  \\ 
  &= T(q_1^2  = x^2 m_1 ,q_2^2  = (p_{12}  - xp_1 )^2 ,q_3^2  = m_3 ) \\ 
 \end{array}
\label{1l3m}
\ee
where the underlying kinematics is defined by $p_1^2=m_1, p_2^2=0, p_{12}^2=(p_1+p_2)^2=m_3$
and 
\[ m_1=\frac{q_1^4-2 q_1^2 q_2^2 +(q_2^2-q_3^2)^2+\lambda (q_1^2-q_2^2+q_3^2)}{2 q_1^2}\]
\[ x=\frac{q_1^2-q_2^2+q_3^2-\lambda}{2 q_3^2}\]
\[ \lambda=\sqrt{q_1^4+q_2^4+q_3^4-2 q_1^2 q_2^2-2 q_2^2 q_3^2-2 q_3^2 q_1^2}\].

The differential equation for $G_{111}$, is given simply by
\be
\frac{\partial }{{\partial x}}G_{111}  =  - \frac{1}{x}G_{111}  + m_1 xG_{121}  + \frac{1}{x}G_{021} 
\ee
 and using IBPI we obtain 
 \[
\begin{array}{ll}
 m_1 xG_{121}  + \frac{1}{x}G_{021} &  = \left( {\frac{1}{{x - 1}} + \frac{1}{{x - m_3 /m_1 }}} \right)\left( {\frac{{d - 4}}{2}} \right)G_{111}  \\ 
  &+ \frac{{d - 3}}{{m_1  - m_3 }}\left( {\frac{1}{{x - 1}} - \frac{1}{{x - m_3 /m_1 }}} \right)\left( {\frac{{G_{101}  - G_{110} }}{x}} \right) \\ 
 \end{array}
\]
The integrating factor $M$ is given by
\[
M=x\left( {1 - x} \right)^{\frac{{4 - d}}{2}} \left( { - m_3  + m_1 x} \right)^{\frac{{4 - d}}{2}} 
\]
 and the DE takes the form, $d=4-2 \varepsilon$,
 \be
\frac{\partial }{{\partial x}}MG_{111}  = c_\Gamma  \frac{1}{\varepsilon }\left( {1 - x} \right)^{ - 1 + \varepsilon } \left( { - m_3  + m_1 x} \right)^{ - 1 + \varepsilon } \left( {\left( { - m_1 x^2 } \right)^{ - \varepsilon }  - \left( { - m_3 } \right)^{ - \varepsilon } } \right)
\ee
where we have used the known expression for the two-point function,
\be
\int {\frac{{d^d k}}{{i\pi ^{d/2} }}\frac{1}{{\left( { - k^2 } \right)\left( { - \left( {k + p} \right)^2 } \right)}}}  = c_\Gamma  \frac{{ - 2}}{{\left( {d - 3} \right)\left( {d - 4} \right)}}\left( { - p^2 } \right)^{d/2 - 2} 
\ee
with
\be
c_\Gamma   = \frac{{\Gamma \left( {3 - d/2} \right)\Gamma \left( {d/2 - 1} \right)^2 }}{{\Gamma \left( {d - 3} \right)}}
\ee
The left-hand side results to 
\be
\int\limits_0^x {dt} \frac{\partial }{{\partial t}}\left( {MG_{111} } \right) = \left( {MG_{111} } \right)_x  - \left( {MG_{111} } \right)_{x = 0} 
\ee
We need therefore, in general, to fix the boundary condition. In the case under consideration we find {\it a posteriori} that $\left( {MG_{111} } \right)_{x = 0}=0$, that 
can easily be justified since $M$ is proportional to $x$ and $G_{111}(x=0)$ is not singular. 
We will see later on how this is justified and how we can address this issue in most general cases when the $x=0$ term is not vanishing.
The right hand side of the equation can now be expanded around  $\varepsilon=0$ and integrated over $x$. All terms can be brought in the form 

\[
\int\limits_0^x {dt} \frac{1}{{t - a_n }}G\left( {a_{n - 1} , \ldots ,a_1 ,t} \right)
\]
by using partial fractioning and the shuffle algebra properties of GPs,
resulting to an expression of the MI in terms of GPs, with argument $x$ and weights $a_i$ that are given in terms of invariants and are independent of $x$. 
The GPs~\cite{Goncharov:1998kja,Goncharov:2001} are defined as follows
\be
G\left( {a_n , \ldots ,a_1 ,x} \right) = \int\limits_0^x {dt} \frac{1}{{t - a_n }}G\left( {a_{n - 1} , \ldots ,a_1 ,t} \right)
\label{gpdef}
\ee
with the special cases, $G(x)=1$ and
\[
G\left( {\underbrace {0, \ldots 0}_n,x} \right) = \frac{1}{{n!}}\log ^n \left( x \right)
\]

This will accomplish the task unless singularities are present that have to be taken properly into account.
These singularities reflect the fact that for certain MI the limits $\varepsilon\to 0$ and $x\to 0 $ or $x \to 1$ do not commute.
For instance, the first non-trivial remark is to observe the existence of a singularity at $x=1$, which of course is regulated by keeping $\varepsilon$ different from zero. In order to properly 
handle the singularity at $x=1$ we have to use the following trivial decomposition

\be
\begin{array}{*{20}c}
   {\int\limits_0^x {dt} \left( {1 - t} \right)^{ - 1 + \varepsilon } f\left( t \right) = \int\limits_0^x {dt} \left( {1 - t} \right)^{ - 1 + \varepsilon } f\left( 1 \right) + \int\limits_0^x {dt} \left( {1 - t} \right)^\varepsilon  \frac{{f(t) - f\left( 1 \right)}}{{\left( {1 - t} \right)}}}  \\
   { = f\left( 1 \right)\frac{{1 - \left( {1 - x} \right)^\varepsilon  }}{\varepsilon } + \int\limits_0^x {dt} \frac{{f(t) - f\left( 1 \right)}}{{\left( {1 - t} \right)}}\left( {1 + \varepsilon G\left( {1,t} \right) + \varepsilon ^2 G\left( {1,1,t} \right) +  \ldots } \right)}  \\
\end{array}
\ee
This procedure essentially allows to properly take the limit $x \to 1$, by simply putting $x_1  \equiv \left( {1 - x} \right)^{ - \varepsilon }  \to 0$.
The result for the 3-mass triangle is given by 
\be
G_{111}  = \frac{{c_\Gamma  }}{{(m_1  - m_3) }x}{\cal I}
\label{3mass}
\ee

{\tiny
\begin{dmath*}
{\cal I}=\frac{-\left(-m_1\right){}^{-\varepsilon }+\left(-m_3\right){}^{-\varepsilon }+\left(\left(-m_1\right){}^{-\varepsilon
   }-\left(-m_3\right){}^{-\varepsilon }\right) x_1}{\varepsilon ^2}+\frac{\left(\left(-m_1\right){}^{-\varepsilon
   }-\left(-m_3\right){}^{-\varepsilon }\right) x_1 G\left(\frac{m_3}{m_1},1\right)-\left(\left(-m_1\right){}^{-\varepsilon
   }-\left(-m_3\right){}^{-\varepsilon }\right)
   \left(G\left(\frac{m_3}{m_1},1\right)-G\left(\frac{m_3}{m_1},x\right)\right)}{\varepsilon }+\left(\left(-m_1\right){}^{-\varepsilon
   }-\left(-m_3\right){}^{-\varepsilon }\right) \left(G\left(\frac{m_3}{m_1},1\right)
   G\left(\frac{m_3}{m_1},x\right)-G\left(\frac{m_3}{m_1},\frac{m_3}{m_1},1\right)-G\left(\frac{m_3}{m_1},\frac{m_3}{m_1}
   ,x\right)\right)
   +x_1 \left(-2 G(0,1,x) \left(-m_1\right){}^{-\varepsilon }+2 G\left(0,\frac{m_3}{m_1},x\right)
   \left(-m_1\right){}^{-\varepsilon }+2 G\left(\frac{m_3}{m_1},1,x\right) \left(-m_1\right){}^{-\varepsilon
   }+G\left(\frac{m_3}{m_1},\frac{m_3}{m_1},1\right) \left(-m_1\right){}^{-\varepsilon }-G\left(\frac{m_3}{m_1},x\right) \log
   (1-x) \left(-m_1\right){}^{-\varepsilon }-2 G\left(\frac{m_3}{m_1},x\right) \log (x) \left(-m_1\right){}^{-\varepsilon }+2 \log (1-x)
   \log (x) \left(-m_1\right){}^{-\varepsilon }-2 \left(-m_3\right){}^{-\varepsilon }
   G\left(\frac{m_3}{m_1},1,x\right)-\left(-m_3\right){}^{-\varepsilon }
   G\left(\frac{m_3}{m_1},\frac{m_3}{m_1},1\right)-\left(\left(-m_1\right){}^{-\varepsilon }-\left(-m_3\right){}^{-\varepsilon
   }\right) G\left(\frac{m_3}{m_1},1\right) \left(G\left(\frac{m_3}{m_1},x\right)-\log (1-x)\right)+\left(-m_3\right){}^{-\varepsilon }
   G\left(\frac{m_3}{m_1},x\right) \log (1-x)\right)
   \\
   +\varepsilon  \left(\left(\left(-m_1\right){}^{-\varepsilon
   }-\left(-m_3\right){}^{-\varepsilon }\right) \left(G\left(\frac{m_3}{m_1},x\right)
   G\left(\frac{m_3}{m_1},\frac{m_3}{m_1},1\right)-G\left(\frac{m_3}{m_1},1\right)
   G\left(\frac{m_3}{m_1},\frac{m_3}{m_1},x\right)-G\left(\frac{m_3}{m_1},\frac{m_3}{m_1},\frac{m_3}{m_1},1\right)+G\left(\frac{m_3}{m_1},\frac{m_3}{m_1},\frac{m_3}{m_1},x\right)\right)
   +\frac{1}{2} x_1 \left(\left(\left(-m_1\right){}^{-\varepsilon }-\left(-m_3\right){}^{-\varepsilon }\right)
   G\left(\frac{m_3}{m_1},1\right) \left(\log ^2(1-x)+2
   G\left(\frac{m_3}{m_1},\frac{m_3}{m_1},x\right)\right)+G\left(\frac{m_3}{m_1},x\right) \left(4 \log ^2(x)-2
   \left(\left(-m_1\right){}^{-\varepsilon }-\left(-m_3\right){}^{-\varepsilon }\right)
   G\left(\frac{m_3}{m_1},\frac{m_3}{m_1},1\right)+2 \left(\left(-m_3\right){}^{-\varepsilon }-\left(-m_1\right){}^{-\varepsilon
   }\right) G\left(\frac{m_3}{m_1},1\right) \log (1-x)\right)+2
   \left(G\left(\frac{m_3}{m_1},\frac{m_3}{m_1},\frac{m_3}{m_1},1\right) \left(-m_1\right){}^{-\varepsilon
   }+G\left(\frac{m_3}{m_1},\frac{m_3}{m_1},1\right) \log (1-x) \left(-m_1\right){}^{-\varepsilon }-2 \log (1-x) \log ^2(x)-4
   G(0,0,1,x)
   \\
   +4 G\left(0,0,\frac{m_3}{m_1},x\right)-2 G(0,1,1,x)+4 G\left(0,\frac{m_3}{m_1},1,x\right)-2
   G\left(0,\frac{m_3}{m_1},\frac{m_3}{m_1},x\right)+2 G\left(\frac{m_3}{m_1},0,1,x\right)
   -2
   G\left(\frac{m_3}{m_1},0,\frac{m_3}{m_1},x\right)
   \\
   -\left(-m_3\right){}^{-\varepsilon }
   G\left(\frac{m_3}{m_1},\frac{m_3}{m_1},\frac{m_3}{m_1},1\right)-\left(-m_3\right){}^{-\varepsilon }
   G\left(\frac{m_3}{m_1},\frac{m_3}{m_1},1\right) \log (1-x)+\log ^2(1-x) \log (x)
   \\
   +4 G(0,1,x) \log (x)-4
   G\left(0,\frac{m_3}{m_1},x\right) \log (x)-4 G\left(\frac{m_3}{m_1},1,x\right) \log (x)+2
   G\left(\frac{m_3}{m_1},\frac{m_3}{m_1},x\right) \log (x)\right)\right)\right)
    \end{dmath*}
}

\begin{figure}
\begin{center}
\includegraphics[width=8cm]{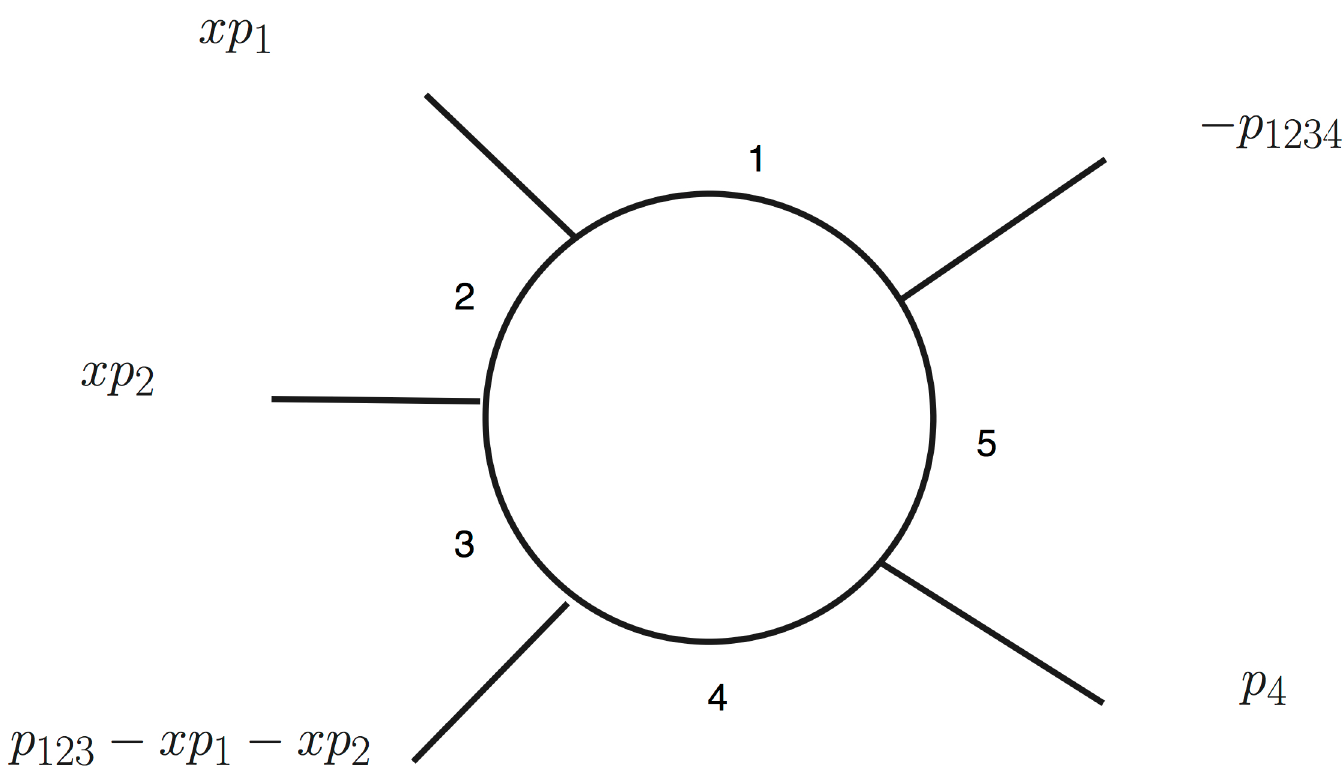}
%\[ \epsfig{figure=pentagon.eps,width=8cm} \]
\caption{The one-loop pentagon graph with one off-shell leg, $G_{11111}$. The labels refer to the denominators in \eqn{1l5p}.}
\label{fig:1l5p}
\end{center}
\end{figure}

It is straightforward to see that by taking now the limit $x\to 1$, as described above, we can easily reproduce the result for the two-mass triangle
\[
I_{2m}  = \frac{{c_\Gamma  }}{{\varepsilon ^2 }}\frac{{\left( {\left( { - m_3 } \right)^{ - \varepsilon }  - \left( { - m_1 } \right)^{ - \varepsilon } } \right)}}{{\left( {m_1  - m_3 } \right)}}
\]
We turn now to a less trivial example, namely the one-loop 5-point MI, Fig.~\ref{fig:1l5p}
\be
\int {\frac{{d^d k}}{{i\pi ^{d/2} }}\frac{1}{{\left( { - k^2 } \right)\left( { - \left( {k + x\,p_1 } \right)^2 } \right)\left( { - \left( {k + xp_{12} } \right)^2 } \right)\left( { - \left( {k + p_{123} } \right)^2 } \right)\left( { - \left( {k + p_{1234} } \right)^2 } \right)}}} 
\label{1l5p}
\ee
with $p_i^2=0,\;\; i=1,\ldots,4$, the obvious notation $p_{i...j}  = p_i  +  \ldots  + p_j $ and $p_{1234}^2=(-p_5)^2=0$.
The correspondence between our kinematics and the one used in Eq.(5.8) of reference~\cite{Bern:1993kr} is given by 

\[
\begin{array}{l}
 s_{12}\equiv(p_1+p_2)^2  \to \frac{{m_5 ^4  - 2m_5 ^2 \left( {s_{12}  + s_{34} } \right) + s_{12} ^2  + s_{34} ^2  - \left( {m_5 ^2  - s_{12}  - s_{34} } \right)\lambda }}{{2s_{34} }} \\ 
 s_{23}\equiv(p_2+p_3)^2  \to \frac{{m_5 ^2 \left( {s_{12}  - s_{45} } \right) - s_{12} ^2  + s_{12} s_{34}  + s_{12} s_{45}  + s_{34} s_{45}  - \left( {s_{12}  - s_{45} } \right)\lambda }}{{2s_{34} }} \\ 
 s_{34}\equiv(p_3+p_4)^2  \to \frac{{m_5 ^4  - m_5 ^2 \left( {2s_{12}  + s_{34}  + s_{51} } \right) + s_{12} ^2  - s_{12} s_{34}  + s_{12} s_{51}  + s_{34} s_{51}  - \left( {m_5 ^2  - s_{12}  - s_{51} } \right)\lambda }}{{2s_{34} }} \\ 
 s_{45}\equiv(p_4+p_5)^2  \to s_{12}  \\ 
 s_{51}\equiv(p_5+p_1)^2  \to \frac{{s_{23} \left( { - m_5 ^2  + s_{12}  + s_{34}  + \lambda } \right)}}{{2s_{34} }} \\ 
 x \to \frac{{ - m_5 ^2  + s_{12}  + s_{34}  - \lambda }}{{2s_{12} }} \\ 
 \lambda  = (s_{12}-s_{45})x \to \sqrt {\left( {s_{12}  + s_{34}  - m_5 ^2 } \right)^2  - 4s_{12} s_{34} }  \\ 
 \end{array}
\]

Notice that we have inserted the parameter dependence in two denominators: the reason is that with only one denominator deformed, the resulting DE exhibits non-rational (square root) dependence on the $x-$parameter
that violates the direct expressibility in terms of GPs. It is also true that, by simple inspection of the possible pinched contributions arising form the original 5-point MI, that the 3-mass sub-topology does not correspond to the parametrization used above, \eqn{3mass},
unless two of the momenta are rescaled.

The DE for the 5-point MI takes the simple form
\be
\frac{\partial }{{\partial x}}G_{11111}  =  - \frac{2}{x}G_{11111}  + s_{12} xG_{11211}  + \frac{1}{x}\left( {G_{02111}  + G_{01211} } \right)
\ee
which after IBPI results to
\be
\begin{array}{l}
 M^{ - 1} \frac{\partial }{{\partial x}}\left( {MG_{11111} } \right) = c_{00110} G_{00110}  + c_{10010} G_{10010}  + c_{10100} G_{10100}  \\ 
 \,\,\,\,\,\,\,\,\,\,\,\,\,\,\,\,\,\,\,\,\,\,\,\,\,\,\,\,\,\,\,\,\,\,\,\,\,\,\,\,\, + c_{10110} G_{10110}  \\ 
 \,\,\,\,\,\,\,\,\,\,\,\,\,\,\,\,\,\,\,\,\,\,\,\,\,\,\,\,\,\,\,\,\,\,\,\,\,\,\,\,\, + c_{10111} G_{10111}  + c_{11011} G_{11011}  + c_{11101} G_{11101}  + c_{11110} G_{11110}  
 \end{array}
 \label{5point}
\ee
where the constants $c_{...}$ are rational functions of $x$. The integrating factor is given by
\be
M = \left( { - s_{23} } \right)^{1 + \varepsilon } \left( { - s_{34} } \right)^{1 + \varepsilon } \left( {1 - \frac{1}{{r_1 }}} \right)^{ - 1 - \varepsilon } \left( {1 - \frac{1}{{r_2 }}} \right)^{ - 1 - \varepsilon } x^2 \left( {1 - \frac{x}{{r_1 }}} \right)^{1 + \varepsilon } \left( {1 - \frac{x}{{r_2 }}} \right)^{1 + \varepsilon } 
\label{m11111}\ee
with
\[
r_1  =\frac{-\sqrt{\Delta}+s_{12} s_{23}-2 s_{12} s_{45}-s_{12} s_{51}-s_{23}
   s_{34}+s_{34} s_{45}-s_{45} s_{51}}{2 s_{12} (s_{23}-s_{45}-s_{51})}
\]
and
\[
r_2  =\frac{\sqrt{\Delta}+s_{12} s_{23}-2 s_{12} s_{45}-s_{12} s_{51}-s_{23}
   s_{34}+s_{34} s_{45}-s_{45} s_{51}}{2 s_{12} (s_{23}-s_{45}-s_{51})}
\]
\[   
\Delta=(-s_{12} s_{23}+2 s_{12} s_{45}+s_{12} s_{51}+s_{23} s_{34}-s_{34}
   s_{45}+s_{45} s_{51})^2+4 s_{12} s_{45} (s_{12}-s_{34}+s_{51})
   (s_{23}-s_{45}-s_{51})
\]
Although a bit more lengthy, the structure of individual terms follows the pattern we expect: for instance the term $M c_{00110} G_{00110}$ is given by
\[
\begin{array}{ll}
 M c_{00110} G_{00110} & = \frac{{c_\Gamma  }}{\varepsilon }\left( {\frac{1}{{s_{12} s_{45} \left( {1 - x} \right)}} + \frac{1}{{s_{45} \left( { - s_{12}  + s_{34}  + s_{12} x} \right)}} - \frac{1}{{s_{45} \left( { - s_{45}  + s_{12} x} \right)}} + \frac{{ - s_{23}  + s_{45} }}{{s_{12} s_{45} \left( { - s_{45}  + \left( { - s_{23}  + s_{45} } \right)x} \right)}}} \right) \\ 
  &\times \left( {1 - \frac{x}{{r_1 }}} \right)^\varepsilon  \left( {1 - \frac{x}{{r_2 }}} \right)^\varepsilon  \left( {1 - x} \right)^{ - \varepsilon } \left( {1 - \frac{{s_{12} }}{{s_{45} }}x} \right)^{ - \varepsilon }  \\ 
  &\times \left( {1 - \frac{1}{{r_1 }}} \right)^{ - \varepsilon } \left( {1 - \frac{1}{{r_2 }}} \right)^{ - \varepsilon } \left( { - s_{23} } \right)^\varepsilon  \left( { - s_{34} } \right)^\varepsilon  \left( { - s_{45} } \right)^{ - \varepsilon } \left( {s_{12}  + s_{45} } \right) \\ 
 \end{array}
\]
which can now be treated in exactly the same way as in the case of the 3-mass triangle.

In the 5-point case of course more MI are involved including also non-IPBI-reducible MI with four external legs.
All these can be treated under the same framework. For instance the $G_{10111}$ two-mass hard one obeys the following DE
\be
\frac{\partial }{{\partial x}}G_{10111}  =  - \frac{1}{x}G_{10111}  + s_{12} xG_{10211}  + \frac{1}{x}G_{00211} 
\ee
with an integrating factor given by
\[
M = x\left( {1 - \frac{{s_{12} }}{{s_{34} }}} \right)^{1 + 2\varepsilon } \left( { - s_{34} } \right)^{1 + 2\varepsilon } \left( {1 - \frac{x}{{1 - \frac{{s_{34} }}{{s_{12} }}}}} \right)^{1 + 2\varepsilon } 
\]
After IBPI the DE takes the form
\[
\begin{array}{ll}
 M^{-1}\frac{\partial }{{\partial x}}\left( {MG_{10111} } \right) &= \frac{{2(1 - 2\varepsilon ) }}{{x^2 \left( {s_{12} (x - 1) + s_{34} } \right)\left( {s_{45} (x - 1) + s_{34} x} \right)}} G_{00101}
+ \frac{{(1 - 2\varepsilon )\left( {s_{45} (x - 2) + s_{12} x} \right) }}{{s_{45} (x - 1)x^2 \left( {s_{12} (x - 1) + s_{34} } \right)\left( {s_{45}  - s_{12} x} \right)}} G_{10100}
\\ 
  &+ \frac{{(1 - 2\varepsilon )\left( {s_{45} \left( {s_{34} (x - 2) + s_{45} (x - 1)} \right) + s_{12} \left( { - s_{45} x + s_{34} x + s_{45} } \right)} \right) }}{{s_{45} (x - 1)x\left( {s_{12} (x - 1) + s_{34} } \right)\left( {s_{12} x - s_{45} } \right)\left( {s_{45} (x - 1) + s_{34} x} \right)}}G_{00110} 
\\
&+ \frac{{(1 - 2\varepsilon )\left( { - 2s_{12} x + s_{12}  + s_{45} } \right) }}{{s_{45} (x - 1)x\left( {s_{12} (x - 1) + s_{34} } \right)\left( {s_{45}  - s_{12} x} \right)}} G_{10010}
\\ 
  &+ \frac{{\varepsilon \left( {s_{12}  - s_{45} } \right)^2  }}{{s_{45} (x - 1)\left( {s_{12} (x - 1) + s_{34} } \right)\left( {s_{45}  - s_{12} x} \right)}} G_{10110}
 \end{array}
\]
fully expressed in terms of $2-$ and $3-$ point functions given by
\be
\begin{array}{l}
G_{00101}  = c_\Gamma  \frac{{ - 2}}{{\left( {d - 4} \right)\left( {d - 3} \right)}}x^{d/2 - 2} \left( {s_{12} \left( {1 - x} \right) - s_{34} } \right)^{d/2 - 2} 
\\
G_{00110}  = c_\Gamma  \frac{{ - 2}}{{\left( {d - 4} \right)\left( {d - 3} \right)}}\left( {1 - x} \right)^{d/2 - 2} \left( { - s_{45}  + s_{12} x} \right)^{d/2 - 2} 
\\
G_{10010}  = c_\Gamma  \frac{{ - 2}}{{\left( {d - 4} \right)\left( {d - 3} \right)}}\left( { - s_{45} } \right)^{d/2 - 2} 
\\
G_{10100}  = c_\Gamma  \frac{{ - 2}}{{\left( {d - 4} \right)\left( {d - 3} \right)}}\left( { - s_{12} } \right)^{d/2 - 2} \left( x \right)^{d - 4} 
   \end{array}
\ee
and $G_{10110}$ given by \eqn{3mass} with $m_3=s_{45}$, $m_1=s_{12}$.

The singularity structure of the right-hand side is now richer. Singularities at $x=0$ are all proportional to $x^{ - 1 - 2\varepsilon } $ and $x^{ - 1 - \varepsilon } $ and can easily be integrated
by the following decomposition 
\be
\begin{array}{l}
 \int\limits_0^x {dt\,\,t^{ - 1 - 2\varepsilon } F\left( t \right)}  = F\left( 0 \right)\int\limits_0^x {dt\,\,t^{ - 1 - 2\varepsilon } }  + \int\limits_0^x {dt\,\,\frac{{F\left( t \right) - F\left( 0 \right)}}{t}} t^{ - 2\varepsilon }  \\ 
  = F\left( 0 \right)\frac{{x^{ - 2\varepsilon } }}{{\left( { - 2\varepsilon } \right)}} + \int\limits_0^x {dt\,\,\frac{{F\left( t \right) - F\left( 0 \right)}}{t}} \left( {1 - 2\varepsilon \log \left( t \right) + 2\varepsilon ^2 \log ^2 \left( t \right) + ...} \right) \\ 
 \end{array}
\ee
As it turns out the first term is very welcome, as it provides the needed boundary term $\left( {MG_{10111} } \right)_{x = 0}$\footnote{In general the $x=0$ limit does not commute with the integration over the loop momentum.}. 
This is quite remarkable: the regulated DE provides its full solution without any need for an independent calculation of 
the boundary term. Moreover this pattern repeats itself in all cases, with the exception of $G_{11101}$ as we will see below.
Singularities proportional to $(1-x)^{ - 1 - \varepsilon } $ can be treated as in the case of the 3-mass triangle. Finally the apparent singularity proportional to the unregulated term $1/(1-x)$, cancels out when all relevant terms in the
right-hand side are combined. 
The result for the two-mass hard box is
\be
G_{10111}  = \frac{{c_\Gamma  }}{{xs_{45} \left( {s_{34}  - s_{12} \left( {1 - x} \right)} \right)}}\sum\limits_{i \ge  - 2} {\varepsilon ^i f_i } 
\ee
{\tiny 
\begin{dmath*}
 f_{ - 2}  = 2-x_1 
\end{dmath*}
\begin{dmath*}
 f_{ - 1}  =  - 4G\left( {1 - \frac{{s_{34} }}{{s_{12} }},x} \right) + G\left( {\frac{{s_{45} }}{{s_{12} }},x} \right) - 4G\left( {\frac{{s_{34} }}{{s_{12} }},1} \right) - G\left( {\frac{{s_{45} }}{{s_{12} }},1} \right) + 2\log \left( { - s_{12} } \right) - 2\log \left( { - s_{34} } \right) - 2\log \left( { - s_{45} } \right) 
  + x_1 \left( {2G\left( {1 - \frac{{s_{34} }}{{s_{12} }},x} \right) + 2G\left( {\frac{{s_{34} }}{{s_{12} }},1} \right) + G\left( {\frac{{s_{45} }}{{s_{12} }},1} \right) - \log \left( { - s_{12} } \right) + 2\log \left( { - s_{45} } \right)} \right) 
\end{dmath*}
\begin{dmath*}
f_0=2 G\left(0,1-\frac{s_{34}}{s_{12}},x\right)-2 G\left(0,\frac{s_{45}}{s_{12}},x\right)+2
   G\left(0,\frac{s_{45}}{s_{34}+s_{45}},x\right)-2 G\left(1-\frac{s_{34}}{s_{12}},1,x\right)+8
   G\left(1-\frac{s_{34}}{s_{12}},1-\frac{s_{34}}{s_{12}},x\right)-2
   G\left(1-\frac{s_{34}}{s_{12}},\frac{s_{45}}{s_{12}},x\right)-G\left(\frac{s_{45}}{s_{12}},\frac{s_{45}}{s_{12}},x\right)+2
   G\left(\frac{s_{45}}{s_{34}+s_{45}},1,x\right)-2 G\left(\frac{s_{45}}{s_{34}+s_{45}},1-\frac{s_{34}}{s_{12}},x\right)+2
   G\left(\frac{s_{45}}{s_{34}+s_{45}},\frac{s_{45}}{s_{12}},x\right)+x_1 \left(-4
   G\left(1-\frac{s_{34}}{s_{12}},1-\frac{s_{34}}{s_{12}},x\right)+2 \left(\log \left(-s_{12}\right)-2 \log
   \left(1-\frac{s_{12}}{s_{34}}\right)-\log \left(1-\frac{s_{12}}{s_{45}}\right)-2 \log \left(-s_{45}\right)\right)
   G\left(1-\frac{s_{34}}{s_{12}},x\right)+2 G\left(0,\frac{s_{45}}{s_{12}},1\right)-2 G(0,1,1)+\frac{1}{2} \left(\log
   ^2\left(-s_{12}\right)-4 \log ^2\left(1-\frac{s_{12}}{s_{34}}\right)-\log ^2\left(1-\frac{s_{12}}{s_{45}}\right)-2 \log
   ^2\left(-s_{45}\right)+\left(4 \log \left(1-\frac{s_{12}}{s_{34}}\right)-2 \log \left(1-\frac{s_{12}}{s_{45}}\right)\right) \log
   \left(-s_{12}\right)-4 \log \left(1-\frac{s_{12}}{s_{34}}\right) \left(\log \left(1-\frac{s_{12}}{s_{45}}\right)+2 \log
   \left(-s_{45}\right)\right)\right)\right)+\left(-4 \log \left(-s_{12}\right)+8 \log \left(1-\frac{s_{12}}{s_{34}}\right)+4 \log
   \left(-s_{34}\right)+2 \log \left(1-\frac{s_{12}}{s_{45}}\right)+4 \log \left(-s_{45}\right)\right)
   G\left(1-\frac{s_{34}}{s_{12}},x\right)-2 \left(\log \left(1-\frac{s_{12}}{s_{34}}\right)+\log \left(-s_{34}\right)-\log
   \left(-s_{45}\right)+\log (x)\right) G\left(\frac{s_{45}}{s_{34}+s_{45}},x\right)+\log (1-x) \left(2
   G\left(1-\frac{s_{34}}{s_{12}},x\right)-G\left(\frac{s_{45}}{s_{12}},x\right)+2 \log \left(1-\frac{s_{12}}{s_{34}}\right)+\log
   \left(1-\frac{s_{12}}{s_{45}}\right)+2 \log (x)\right)+\left(\log \left(-s_{12}\right)-2 \log \left(-s_{45}\right)+2 \log
   (x)\right) G\left(\frac{s_{45}}{s_{12}},x\right)-2 G\left(0,\frac{s_{45}}{s_{12}},1\right)-2 G(0,1,x)+2 G(0,1,1)+\log
   \left(-s_{12}\right) \left(-2 \log \left(1-\frac{s_{12}}{s_{34}}\right)+\log \left(1-\frac{s_{12}}{s_{45}}\right)-2 \log
   (x)\right)+2 \log \left(-s_{34}\right) \log (x)+2 \log \left(1-\frac{s_{12}}{s_{34}}\right) \left(\log \left(-s_{34}\right)+\log
   \left(1-\frac{s_{12}}{s_{45}}\right)+2 \log \left(-s_{45}\right)+\log (x)\right)-\log ^2\left(-s_{12}\right)+3 \log
   ^2\left(1-\frac{s_{12}}{s_{34}}\right)+\log ^2\left(-s_{34}\right)+\frac{1}{2} \log ^2\left(1-\frac{s_{12}}{s_{45}}\right)+\log
   ^2\left(-s_{45}\right)-\log ^2(x)
\end{dmath*}
}   
The full result up to order $\varepsilon$ is presented as a Mathematica output in the file \verb+f10111.txt+, attached to this paper. 
The nice property of the above expressions is that, by putting $x_1=0$ and $x=1$, one can smoothly derive the one-mass box to the same order in $\varepsilon$.
Finally the other two-mass hard box, $G_{11110}$, is worked out in exactly the same way.

The one-mass box  $G_{11011}$, is particularly simple and is given by ($x_0\equiv x^{-\varepsilon}$)
\be
G_{11011}  = \frac{{2c_\Gamma  }}{{xs_{45} s_{51} }}\sum\limits_{i \ge  - 2} {\varepsilon ^i f_i } 
\ee
with
{\tiny
\begin{dgroup*}
\begin{dmath*}
f_{ - 2}  = x_0 \left( { - s_{51} } \right)^{ - \varepsilon } 
\end{dmath*}
\begin{dmath*}
f_{ - 1}  = \left( { - s_{45} } \right)^{ - \varepsilon } G\left( {\frac{{s_{45} }}{{s_{45}  - s_{23} }},x} \right) + \left( {x_0 \left( { - s_{51} } \right)^{ - \varepsilon }  - \left( { - s_{45} } \right)^{ - \varepsilon } } \right)G\left( {\frac{{s_{45} }}{{ - s_{23}  + s_{45}  + s_{51} }},x} \right)
\end{dmath*}
\begin{dmath*}
f_0  = \left( { - s_{45} } \right)^{ - \varepsilon } \left( { - G\left( {\frac{{s_{45} }}{{s_{45}  - s_{23} }},\frac{{s_{45} }}{{s_{45}  - s_{23} }},x} \right)} \right) + \left( { - s_{45} } \right)^{ - \varepsilon } G\left( {\frac{{s_{45} }}{{ - s_{23}  + s_{45}  + s_{51} }},\frac{{s_{45} }}{{s_{45}  - s_{23} }},x} \right) 
  + x_0 \left( { - s_{51} } \right)^{ - \varepsilon } G\left( {0,\frac{{s_{45} }}{{ - s_{23}  + s_{45}  + s_{51} }},x} \right) 
\end{dmath*}
\begin{dmath*}
 f_1  = \left( { - s_{45} } \right)^{ - \varepsilon } G\left( {\frac{{s_{45} }}{{s_{45}  - s_{23} }},\frac{{s_{45} }}{{s_{45}  - s_{23} }},\frac{{s_{45} }}{{s_{45}  - s_{23} }},x} \right) - \left( { - s_{45} } \right)^{ - \varepsilon } G\left( {\frac{{s_{45} }}{{ - s_{23}  + s_{45}  + s_{51} }},\frac{{s_{45} }}{{s_{45}  - s_{23} }},\frac{{s_{45} }}{{s_{45}  - s_{23} }},x} \right) \\ 
  + x_0 \left( { - s_{51} } \right)^{ - \varepsilon } G\left( {0,0,\frac{{s_{45} }}{{ - s_{23}  + s_{45}  + s_{51} }},x} \right) 
\end{dmath*}
\end{dgroup*}
 }
Notice that the DE for $G_{11011}$ is also producing the full answer without reference to the boundary term.

Finally the other one-mass box $G_{11101}$  satisfies the following DE
{\footnotesize
\be
\frac{\partial }{{\partial x}}\left( {MG_{11101} } \right) =  - \frac{{2c_\Gamma  }}{{\left( {s_{12} (1 - x) - s_{34}  + s_{51} } \right)}}\left( {\frac{{\left( { - s_{51} } \right)^{ - \varepsilon } x^\varepsilon  }}{{\varepsilon s_{51} }} + \frac{{\left( {1 - \frac{{s_{12} }}{{s_{34} }}} \right)^{ - \varepsilon } \left( { - s_{34} } \right)^{ - \varepsilon } x^\varepsilon  \left( {1 - \frac{{s_{12} x}}{{s_{12}  - s_{34} }}} \right)^{ - \varepsilon } }}{{\varepsilon \left( {s_{12} x - s_{12}  + s_{34} } \right)}}} \right)
\ee
}
with $M=x^{3+2\varepsilon}$. The singularity structure at $x=0$ prohibits us from deriving the full answer without an independent calculation of the boundary term. Nevertheless, this term can easily be derived from $G_{11011}$ box under the proper 
replacements.  
The result is given by
\be
G_{11101}  = \frac{{2c_\Gamma  }}{{x^3 s_{12} s_{51} }}\sum\limits_{i \ge  - 2} {\varepsilon ^i f_i } 
\ee
{\footnotesize
\[
\begin{array}{l}
f_{ - 2}  = x_0 ^2 \left( { - s_{12} } \right)^{ - \varepsilon } \\
 f_{ - 1}  = x_0 ^2 \left( { - s_{12} } \right)^{ - \varepsilon } G\left( {\frac{{s_{51} }}{{s_{12}  - s_{34}  + s_{51} }},1} \right) + x_0 \left( {1 - \frac{{s_{12} }}{{s_{34} }}} \right)^{ - \varepsilon } \left( { - s_{34} } \right)^{ - \varepsilon } G\left( {1 - \frac{{s_{34} }}{{s_{12} }},x} \right) \\ 
  + x_0 \left( {\left( { - s_{51} } \right)^{ - \varepsilon }  - \left( {1 - \frac{{s_{12} }}{{s_{34} }}} \right)^{ - \varepsilon } \left( { - s_{34} } \right)^{ - \varepsilon } } \right)G\left( {\frac{{s_{12}  - s_{34}  + s_{51} }}{{s_{12} }},x} \right) \\ 
f_0  = x_0 ^2 \left( { - s_{12} } \right)^{ - \varepsilon } G\left( {0,\frac{{s_{51} }}{{s_{12}  - s_{34}  + s_{51} }},1} \right) - x_0 \left( {1 - \frac{{s_{12} }}{{s_{34} }}} \right)^{ - \varepsilon } \left( { - s_{34} } \right)^{ - \varepsilon } G\left( {0,1 - \frac{{s_{34} }}{{s_{12} }},x} \right) \\ 
  + x_0 \left( {\left( {1 - \frac{{s_{12} }}{{s_{34} }}} \right)^{ - \varepsilon } \left( { - s_{34} } \right)^{ - \varepsilon }  - \left( { - s_{51} } \right)^{ - \varepsilon } } \right)G\left( {0,\frac{{s_{12}  - s_{34}  + s_{51} }}{{s_{12} }},x} \right) \\ 
  - x_0 \left( {1 - \frac{{s_{12} }}{{s_{34} }}} \right)^{ - \varepsilon } \left( { - s_{34} } \right)^{ - \varepsilon } G\left( {1 - \frac{{s_{34} }}{{s_{12} }},1 - \frac{{s_{34} }}{{s_{12} }},x} \right) + x_0 \left( {1 - \frac{{s_{12} }}{{s_{34} }}} \right)^{ - \varepsilon } \left( { - s_{34} } \right)^{ - \varepsilon } G\left( {\frac{{s_{12}  - s_{34}  + s_{51} }}{{s_{12} }},1 - \frac{{s_{34} }}{{s_{12} }},x} \right) \\ 
 f_1  = x_0 ^2 \left( { - s_{12} } \right)^{ - \varepsilon } G\left( {0,0,\frac{{s_{51} }}{{s_{12}  - s_{34}  + s_{51} }},1} \right) + x_0 G\left( {0,0,1 - \frac{{s_{34} }}{{s_{12} }},x} \right) + x_0 G\left( {0,1 - \frac{{s_{34} }}{{s_{12} }},1 - \frac{{s_{34} }}{{s_{12} }},x} \right) \\ 
  - x_0 G\left( {0,\frac{{s_{12}  - s_{34}  + s_{51} }}{{s_{12} }},1 - \frac{{s_{34} }}{{s_{12} }},x} \right) + x_0 G\left( {1 - \frac{{s_{34} }}{{s_{12} }},1 - \frac{{s_{34} }}{{s_{12} }},1 - \frac{{s_{34} }}{{s_{12} }},x} \right) \\ 
  - x_0 G\left( {\frac{{s_{12}  - s_{34}  + s_{51} }}{{s_{12} }},1 - \frac{{s_{34} }}{{s_{12} }},1 - \frac{{s_{34} }}{{s_{12} }},x} \right) 
 \end{array}
\]
}
In fact in the above result the terms proportional to $x_0^2$ are not provided by the DE itself. 

Having now all the ingredients for the pentagon DE equation, \eqn{5point}, we can proceed to get its solution. 
As in most cases there is no need for an independent evaluation of the boundary term. All singularities at $x=0$ and $x=1$ are treated as before.
In fact there are also singularities proportional to $x^{-2-2\varepsilon}$, that can be handled with the following decomposition
\be
\begin{array}{l}
 \int\limits_0^x {dt} \,\,t^{ - 2 - 2\varepsilon } f\left( t \right) = f\left( 0 \right)\int\limits_0^x {dt} \,\,t^{ - 2 - 2\varepsilon }  + f'\left( 0 \right)\int\limits_0^x {dt} \,\,t^{ - 1 - 2\varepsilon }  \\ 
  + \int\limits_0^x {dt} \,\,\frac{{f\left( t \right) - f\left( 0 \right) - tf'\left( 0 \right)}}{{t^2 }}\left( {1 - 2\varepsilon G\left( {0,t} \right) + 4\varepsilon ^2 G\left( {0,0,t} \right) +  \ldots } \right) \\ 
 \end{array}
\ee
As is readily can be seen from the last terms, new integrals of GPs are needed. In fact we need, in general, integrals of the form
\[
\int\limits_0^x {dt} \left\{ {\frac{1}{{\left( {t - a_n } \right)^2 }},\frac{1}{{t^2 }},1} \right\}G\left( {a_{n - 1} , \ldots ,a_1 ,t} \right)
\]
%Expressions for those integrals are provided in the Appendix \ref{appendix}.
that are easily obtained by integration by parts.

The result for the one-mass pentagon up to order $\varepsilon$, is given by
\be
G_{11111} (x) = \frac{{c_\Gamma  }}{{x^2 s_{23} s_{34} s_{45} }}\left( {1 - \frac{1}{{r_1 }}} \right)\left( {1 - \frac{1}{{r_2 }}} \right)\left( {1 - \frac{x}{{r_1 }}} \right)^{ - 1 - \varepsilon } \left( {1 - \frac{x}{{r_2 }}} \right)^{ - 1 - \varepsilon } \sum\limits_{i \ge  - 2} {\varepsilon ^i f_i }
\ee
where $f_i$ are given in file \verb+full.txt+.
Taking the limit $x\to 1$ ($x_1\to0$) from the previous expression, we get the result for the on-shell pentagon up to order $\varepsilon$, that is given by
\be
G_{11111} (1) = \frac{{c_\Gamma  }}{{s_{12} s_{23} s_{34} s_{45} s_{51} }}\sum\limits_{i \ge  - 2} {\varepsilon ^i f_i }
\ee
where $f_i$ are given in file \verb+full1.txt+. In both files \verb+DD+ denotes $\sqrt{\Delta}$ given above.

\subsection{Two-loop results}

There is no fundamental difference in applying the method at two loops. 
We will see in the sequel that all main aspects remain the same. 
One slight complication is that, in most of the cases, we have to solve a system of coupled DE instead of a single DE, as at one loop. 
This reflects the known structure of the two-loop MI, which is not restricted to scalar Feynman Integrals, in contract to the 
one-loop case. In all cases we have studied, the system of coupled DE has a straightforward solution in terms of expansion in $\varepsilon$~\cite{Henn:2013pwa}. 

\begin{figure}
\begin{center}
\includegraphics[width=5cm]{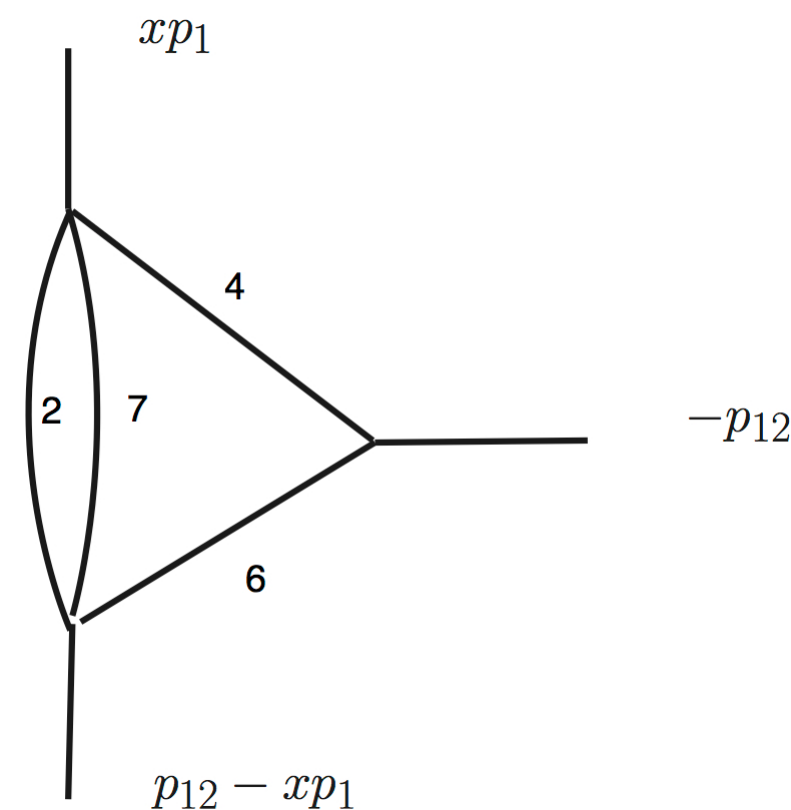}
%\[ \epsfig{figure=3m.eps,,width=5cm} \]
\end{center}
\caption{Two-loop MI with three off-shell legs, $G_{0101011}$: labels are used to identify the propagators, for instance label $2$ refer to $-(k_1+x p_1)^2$, \eqn{3mtop} ($i$ for $a_i=1$).}
\label{fig:2l3m}
\end{figure}

As a first example we will study the triangle with 3 off-shell legs~\cite{Birthwright:2004kk,Chavez:2012kn} given in Fig.~\ref{fig:2l3m}
It belongs to the topology defined by the following integrals,
{\footnotesize
\be
\begin{array}{ll}
 G_{a_1 a_2  \ldots a_7 }  & = \int {\frac{{d^d k_1 }}{{i\pi ^{d/2} }}\frac{{d^d k_2 }}{{i\pi ^{d/2} }}} \frac{1}{{\left( { - k_1^2 } \right)^{a_1 } \left( { - \left( {k_1  + \,x\,p_1 } \right)^2 } \right)^{a_2 } \left( { - \left( {k_1  + p_1  + p_2 } \right)^2 } \right)^{a_3 } }} \\ 
  & \times \frac{1}{{\left( { - k_2^2 } \right)^{a_4 } \left( { - \left( {k_2  - \,x\,p_1 } \right)^2 } \right)^{a_5 } \left( { - \left( {k_2  - p_1  - p_2 } \right)^2 } \right)^{a_6 } \left( { - \left( {k_1  + k_2 } \right)^2 } \right)^{a_7 } }} \\ 
 \end{array}
 \label{3mtop}
\ee
}
where $p_1^2=m_1$, $p_2^2=0$ and $(p_1+p_2)^2=m_3$.

We are interested in $G_{0101011}$. The DE involves also the MI $G_{0201011}$, so we have a system of two coupled DE, as follows:
\[
\begin{array}{ll}
 \frac{\partial }{{\partial x}}\left(M_{0101011} G_{0101011}\right) & = \frac{{A_3 (2 - 3\varepsilon )(1 - x)^{ - 2\varepsilon } x^{\varepsilon  - 1} \left( {m_1 x - m_3 } \right)^{ - 2\varepsilon } }}{{2\varepsilon (2\varepsilon  - 1)}} 
\\ 
  &+ \frac{{m_1 \varepsilon (1 - x)^{ - 2\varepsilon } \left( {m_1 x - m_3 } \right)^{ - 2\varepsilon } }}{{2\varepsilon  - 1}}g(x) 
 \end{array}
\]
\[
\begin{array}{ll}
 \frac{\partial }{{\partial x}}\left(M_{0201011} G_{0201011}\right) &= \frac{{A_3 (3\varepsilon  - 2)(3\varepsilon  - 1)\left( { - m_1 } \right)^{ - 2\varepsilon } (1 - x)^{2\varepsilon  - 1} x^{ - 3\varepsilon } \left( {m_1 x - m_3 } \right)^{2\varepsilon  - 1} }}{{2\varepsilon ^2 }} 
\\ 
  &+ (2\varepsilon  - 1)(3\varepsilon  - 1)(1 - x)^{2\varepsilon  - 1} \left( {m_1 x - m_3 } \right)^{2\varepsilon  - 1} f(x) 
 \end{array}
\]
where $f\left( x \right) \equiv M_{0101011} G_{0101011} $ and $g\left( x \right) \equiv M_{0201011} G_{0201011}$,
$
M_{0201011}  = (1 - x)^{2\varepsilon } x^{\varepsilon  + 1} \left( {m_1 x - m_3 } \right)^{2\varepsilon } 
$
and
$
M_{0101011}  = x^\varepsilon  
$, and
\[
A_3  =  - \frac{{ \Gamma (5 - d)\Gamma \left( {\frac{d}{2} - 1} \right)^3 }}{{\Gamma \left( {\frac{{3d}}{2} - 3} \right)}}
\]
As in the one-loop case, the integration is straightforward and the result is obtained without any reference to the $x=0$ boundary condition, suggesting of course that the latter is vanishing.
More specifically we obtain:
 {\tiny 
  \begin{dmath*}
 f(x) = A_3 \left( { - \frac{1}{{\varepsilon ^2 }} + \frac{{2\log \left( { - m_3 } \right) - \log (x) - \frac{1}{2}}}{\varepsilon }} \right. +  \\ 
 2G\left( {0,\frac{{m_3 }}{{m_1 }},x} \right) - \frac{{4m_1 (x - 1)G(1,0,x)}}{{m_1  - m_3 }} + \frac{{4\left( {m_1 x - m_3 } \right)G\left( {\frac{{m_3 }}{{m_1 }},0,x} \right)}}{{m_1  - m_3 }} -  \\ 
 \frac{{2m_1 (x - 1)\left( {\log \left( { - m_1 } \right) - \log \left( { - m_3 } \right)} \right)G(1,x)}}{{m_1  - m_3 }} + \frac{{2\left( {m_1 x - m_3 } \right)\left( {\log \left( { - m_1 } \right) - \log \left( { - m_3 } \right)} \right)G\left( {\frac{{m_3 }}{{m_1 }},x} \right)}}{{m_1  - m_3 }} +  \\ 
 \left. {2G(0,1,x) + \frac{1}{2}\left( {4\log \left( { - m_3 } \right)\log (x) - 4\log ^2 \left( { - m_3 } \right) + 2\log \left( { - m_3 } \right) - \log ^2 (x) - \log (x) - 2} \right)} \right) \\ 
  \end{dmath*}
  \begin{dmath*}
 g\left( x \right) = 
 \frac{{A_3 }}{\varepsilon }
 \left( {\frac{{4G(1,0,x)}}{{m_1  - m_3 }} - \frac{{4G\left( {\frac{{m_3 }}{{m_1 }},0,x} \right)}}{{m_1  - m_3 }} 
 + \frac{{2\left( {\log \left( { - m_1 } \right) - \log \left( { - m_3 } \right)} \right)G(1,x)}}{{m_1  - m_3 }} 
 + \frac{{2\left( {\log \left( { - m_3 } \right) - \log \left( { - m_1 } \right)} \right)G\left( {\frac{{m_3 }}{{m_1 }},x} \right)}}{{m_1  - m_3 }}} \right) 
  + A_3
 \left(  
 -\frac{8 G(1,0,0,x)}{m_1-m_3}+\frac{6 G(1,0,1,x)}{m_1-m_3}+\frac{6 G\left(1,0,\frac{m_3}{m_1},x\right)}{m_1-m_3}+\frac{8 G(1,1,0,x)}{m_1-m_3}+\frac{4
   G\left(1,\frac{m_3}{m_1},0,x\right)}{m_1-m_3}+\frac{8 G\left(\frac{m_3}{m_1},0,0,x\right)}{m_1-m_3}-\frac{6
   G\left(\frac{m_3}{m_1},0,1,x\right)}{m_1-m_3}-\frac{6 G\left(\frac{m_3}{m_1},0,\frac{m_3}{m_1},x\right)}{m_1-m_3}-\frac{4
   G\left(\frac{m_3}{m_1},1,0,x\right)}{m_1-m_3}-\frac{8 G\left(\frac{m_3}{m_1},\frac{m_3}{m_1},0,x\right)}{m_1-m_3}+\frac{\left(2 \log
   ^2\left(-m_1\right)+\left(9-4 \log \left(-m_3\right)\right) \log \left(-m_1\right)+\log \left(-m_3\right) \left(2 \log
   \left(-m_3\right)-9\right)\right) G\left(\frac{m_3}{m_1},x\right)}{m_1-m_3}+\frac{\left(-2 \log ^2\left(-m_1\right)+\left(4 \log
   \left(-m_3\right)-9\right) \log \left(-m_1\right)+\left(9-2 \log \left(-m_3\right)\right) \log \left(-m_3\right)\right) G(1,x)}{m_1-m_3}+\frac{4
   \left(\log \left(-m_1\right)-\log \left(-m_3\right)\right) G(1,1,x)}{m_1-m_3}+\frac{2 \left(\log \left(-m_1\right)-\log \left(-m_3\right)\right)
   G\left(1,\frac{m_3}{m_1},x\right)}{m_1-m_3}-\frac{2 \left(\log \left(-m_1\right)-\log \left(-m_3\right)\right)
   G\left(\frac{m_3}{m_1},1,x\right)}{m_1-m_3}-\frac{4 \left(\log \left(-m_1\right)-\log \left(-m_3\right)\right)
   G\left(\frac{m_3}{m_1},\frac{m_3}{m_1},x\right)}{m_1-m_3}-\frac{6 \left(\log \left(-m_1\right)-\log \left(-m_3\right)+3\right)
   G(1,0,x)}{m_1-m_3}+\frac{6 \left(\log \left(-m_1\right)-\log \left(-m_3\right)+3\right) G\left(\frac{m_3}{m_1},0,x\right)}{m_1-m_3}
   \right)
   \end{dmath*}
}

\begin{figure}
\begin{center}
\includegraphics[width=8cm]{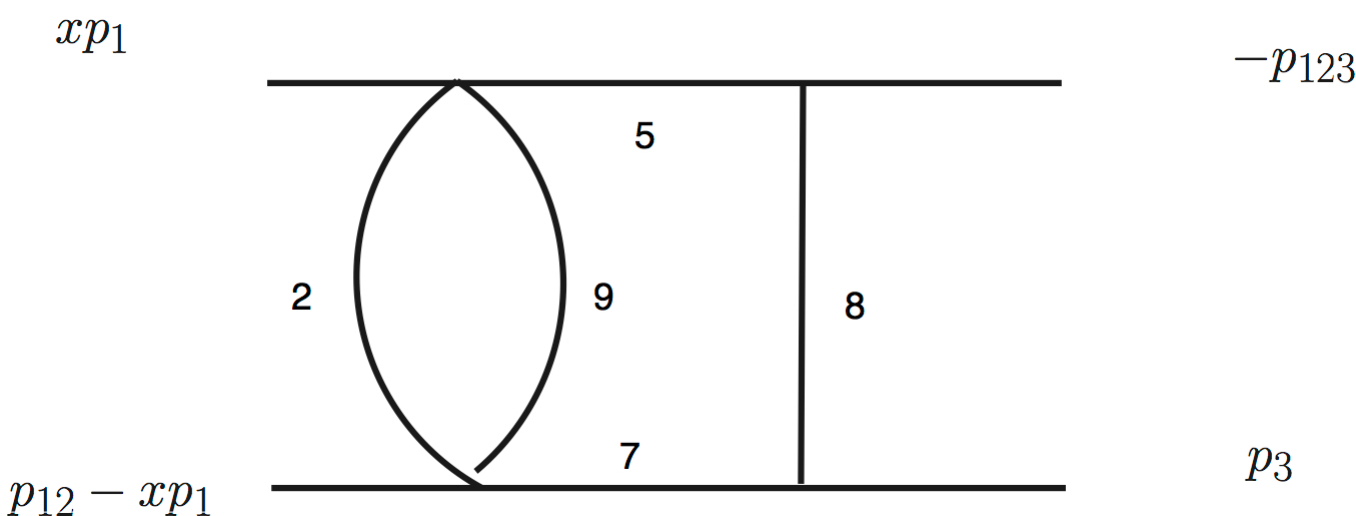}
%\[ \epsfig{figure=box2pie.eps,,width=8cm} \]
\end{center}
\caption{Two-loop box MI with two off-shell legs: the easy one, $G_{010010111}$. See in the text for more details. Labels as in Fig.~\ref{fig:2l3m} with respect to \eqn{box2pie}.} 
\label{fig:box2pie}
\end{figure}

We now turn to 2-loop box graphs. We start with the box with two off-shell legs given by, see Fig.~\ref{fig:box2pie},
\be
\begin{array}{*{20}c}
   {G_{a_1 a_2  \ldots a_9 }  = } & {\int {\frac{{d^d k_1 }}{{i\pi ^{d/2} }}\frac{{d^d k_2 }}{{i\pi ^{d/2} }}} \frac{1}{{\left( { - k_1^2 } \right)^{a_1 } \left( { - \left( {k_1  + \,x\,p_1 } \right)^2 } \right)^{a_2 } \left( { - \left( {k_1  + p_1  + p_2 } \right)^2 } \right)^{a_3 } \left( { - \left( {k_1  + p_1  + p_2  + p_3 } \right)^2 } \right)^{a_4 } }}}  \\
   {} & { \times \frac{1}{{\left( { - k_2^2 } \right)^{a_5 } \left( { - \left( {k_2  - \,p_1 } \right)^2 } \right)^{a_6 } \left( { - \left( {k_2  - p_1  - p_2 } \right)^2 } \right)^{a_7 } \left( { - \left( {k_2  - p_1  - p_2  - p_3 } \right)^2 } \right)^{a_8 } \left( { - \left( {k_1  + k_2 } \right)^2 } \right)^{a_9 } }}} 
\end{array}
\label{box2pie}
\ee
with $p_1^2=0$, $p_2^2=0$, $p_3^2=0$, $s_{ij}=(p_i+p_j)^2$ and $p_{123}^2=q$.

In terms of the conventional kinematics defined as $q_1=x p_1$, $q_2=p_1+p_2-x p_1$, $q_3=p_3$, $q_4=-p_1-p_2-p_3$, $S_{12}=(q_1+q_2)^2$, $S_{23}=(q_2+q_3)^2$, $M_2=q_2^2$ and $M_4=q_4^2$ we
have $x=1-M_2/S_{12}$ and $s_{23}=\left( M_2 M_4-S_{12}S_{23}\right)/\left( M_2-S_{12}\right)$.

The DE turns out to be quite simple
\be
\begin{array}{ll}
 \frac{\partial }{{\partial x}}\left( {x^\varepsilon  G_{010010111} } \right) &=  - \frac{{\left( {9\varepsilon ^2  - 9\varepsilon  + 2} \right)x^{\varepsilon  - 1} }}{{\varepsilon \left( {q(x - 1) - s_{23} x} \right)\left( {q(x - 1) + s_{12} ( - x) - s_{23} x + s_{12} } \right)}} G_{010000011} 
\\ 
  &- \frac{{(3\varepsilon  - 2)(3\varepsilon  - 1) x^{\varepsilon  - 1} }}{{s_{12} (x - 1)\varepsilon \left( {q( - x) + q + s_{12} (x - 1) + s_{23} x} \right)}} G_{010000101} 
 \end{array}
\ee

The existence of a regularized singularity at $x=0$ is treated as in the one-loop case, making the independent evaluation of the $x=0$ boundary condition unnecessary. 
The result is given by 
{\tiny
\begin{dmath}
(q-s_{12})x^\varepsilon G_{010010111}=-\frac{2A_3 \log \left(\frac{q}{s_{12}}\right)}{\varepsilon ^2}+\frac{A_3}{\varepsilon}
\left(-2 G\left(0,\frac{q}{q-s_{23}},x\right)-2 G\left(\frac{s_{12}-q}{-q+s_{12}+s_{23}},1,x\right)+2
   G\left(\frac{s_{12}-q}{-q+s_{12}+s_{23}},\frac{q}{q-s_{23}},x\right)+\log \left(\frac{q}{s_{12}}\right) \left(2
   G\left(\frac{s_{12}-q}{-q+s_{12}+s_{23}},x\right)+2 \log \left(\frac{q}{s_{12}}\right)+4 \log \left(-s_{12}\right)+5\right)+2 G(0,1,x)\right)
   +\frac{A_3}{3} \left(6 G\left(0,0,\frac{q}{q-s_{23}},x\right)+12 G\left(0,\frac{q}{q-s_{23}},\frac{q}{q-s_{23}},x\right)+6
   G\left(0,\frac{s_{12}-q}{-q+s_{12}+s_{23}},1,x\right)-6 G\left(0,\frac{s_{12}-q}{-q+s_{12}+s_{23}},\frac{q}{q-s_{23}},x\right)+12
   G\left(\frac{s_{12}-q}{-q+s_{12}+s_{23}},1,1,x\right)-12 G\left(\frac{s_{12}-q}{-q+s_{12}+s_{23}},\frac{q}{q-s_{23}},\frac{q}{q-s_{23}},x\right)-6
   \log \left(\frac{q}{s_{12}}\right) G\left(0,\frac{s_{12}-q}{-q+s_{12}+s_{23}},x\right)-6 G(0,x)
   \left(G\left(0,\frac{q}{q-s_{23}},x\right)+G\left(\frac{s_{12}-q}{-q+s_{12}+s_{23}},1,x\right)-G\left(\frac{s_{12}-q}{-q+s_{12}+s_{23}},\frac{q}{q-s
   _{23}},x\right)-\log \left(\frac{q}{s_{12}}\right) G\left(\frac{s_{12}-q}{-q+s_{12}+s_{23}},x\right)\right)+3 \left(4 \log \left(-s_{12}\right)+2
   \log (x)+5\right) G\left(\frac{s_{12}-q}{-q+s_{12}+s_{23}},1,x\right)-3 \log \left(\frac{q}{s_{12}}\right) \left(2 \log
   \left(\frac{q}{s_{12}}\right)+4 \log \left(-s_{12}\right)+2 \log (x)+5\right) G\left(\frac{s_{12}-q}{-q+s_{12}+s_{23}},x\right)+3 \left(4 \left(\log
   \left(\frac{q}{s_{12}}\right)+\log \left(-s_{12}\right)\right)+2 \log (x)+5\right) G\left(0,\frac{q}{q-s_{23}},x\right)-3 \left(4 \left(\log
   \left(\frac{q}{s_{12}}\right)+\log \left(-s_{12}\right)\right)+2 \log (x)+5\right)
   G\left(\frac{s_{12}-q}{-q+s_{12}+s_{23}},\frac{q}{q-s_{23}},x\right)+3 G(0,1,x) \left(2 G(0,x)-4 \log \left(-s_{12}\right)-2 \log (x)-5\right)-6
   G(0,0,1,x)-12 G(0,1,1,x)-\log \left(\frac{q}{s_{12}}\right) \left(4 \log ^2\left(\frac{q}{s_{12}}\right)+3 \left(4 \log
   \left(-s_{12}\right)+5\right) \log \left(\frac{q}{s_{12}}\right)+6 \log \left(-s_{12}\right) \left(2 \log
   \left(-s_{12}\right)+5\right)-3\right)\right)
\end{dmath}
}

\begin{figure}
\begin{center}
\includegraphics[width=8cm]{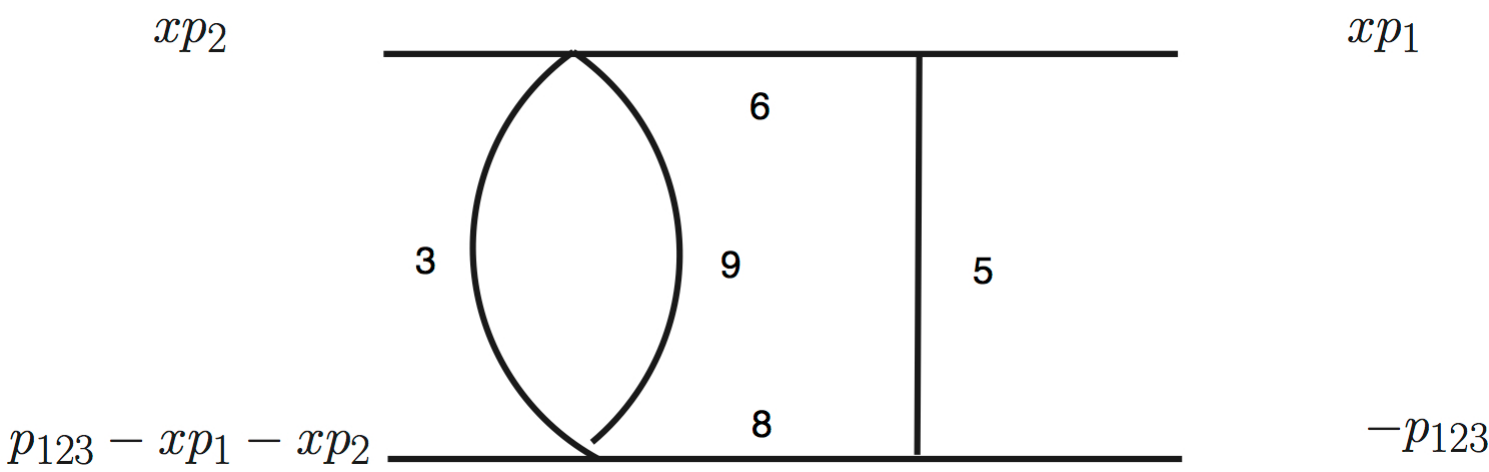}
%\[ \epsfig{figure=box2pih.eps,,width=8cm} \]
\caption{Two-loop box MI with two off-shell legs: the hard one, $G_{001011011}$. See in the text for more details. Labels as in Fig.~\ref{fig:2l3m} with respect to \eqn{box2pih}.} 
\label{fig:box2pih}
\end{center}
\end{figure}

The other box with two off-shell legs, coded as the hard one\footnote{I would like to thank C.~Anastasiou, C.~Duhr and  F.~Chavez for communicating  their preliminary results on the two-loop 
box MI given in Fig.~\ref{fig:box2pih}.}, is given in Fig.~\ref{fig:box2pih}.
It is part of the following topology 
\be
\begin{array}{ll}
 G_{a_1 a_2  \ldots a_9 }  = & \int {\frac{{d^d k_1 }}{{i\pi ^{d/2} }}\frac{{d^d k_2 }}{{i\pi ^{d/2} }}} \frac{1}{{\left( { - k_1^2 } \right)^{a_1 } \left( { - \left( {k_1  + \,x\,p_1 } \right)^2 } \right)^{a_2 } \left( { - \left( {k_1  + x\,p_1  + x\,p_2 } \right)^2 } \right)^{a_3 } \left( { - \left( {k_1  + p_1  + p_2  + p_3 } \right)^2 } \right)^{a_4 } }} \\ 
  & \times \frac{1}{{\left( { - k_2^2 } \right)^{a_5 } \left( { - \left( {k_2  - \,x\,p_1 } \right)^2 } \right)^{a_6 } \left( { - \left( {k_2  - p_1  - p_2 } \right)^2 } \right)^{a_7 } \left( { - \left( {k_2  - p_1  - p_2  - p_3 } \right)^2 } \right)^{a_8 } \left( { - \left( {k_1  + k_2 } \right)^2 } \right)^{a_9 } }} \\ 
 \end{array}
 \label{box2pih}
\ee
with $p_1^2=0$, $p_2^2=0$, $p_3^2=0$ and $p_{123}^2=q$ as before. Notice that the same phenomenon appears as in the one-loop pentagon, namely we have to $x-$parametrize two of the external massless momenta in order to achieve
DE free of square-root terms. This is again related to the 3-off-shell-legs triangle parametrization, now at two loops.  

In terms of the conventional kinematics defined as $q_1=x p_1$, $q_2=x p_2$, $q_3=p_1+p_2+p_3-x p_1-x p_2$, $q_4=-p_1-p_2-p_3$, $S_{12}=(q_1+q_2)^2$, $S_{23}=(q_2+q_3)^2$, $M_3=q_3^2$ and $M_4=q_4^2$ we
have
\[
\begin{array}{l}
s_{12}=\frac{{S_{12}^2+ \left( {M_{4}  - M_{3} } \right)^2 - 2 S_{12} M_3  + \left( {S_{12}  + M_{4}  - M_{3} } \right)\lambda }}{{2 S_{12} }} \\ 
s_{23}=\frac{{S_{12}S_{23}-M_4^2+ M_4 \left( {S_{12}+S_{23}  + M_{3} } \right) - S_{23} M_3  + \left( {S_{23}  - M_{4}  } \right)\lambda }}{{2 S_{12} }} \\ 
x =\frac{{ S_{12}+M_4-M_3   - \lambda }}{{2M_{4} }}, \,\,\,\, q =M_4,  \,\,\,\lambda  = (s_{12}-q)x = \sqrt {\left( {S_{12}  + M_4  - M_3 } \right)^2  - 4 S_{12} M_4 }.  \\ 
 \end{array}
\]

The integrating factor reads
\[
M_{001011011}  = ( - q)^{3\varepsilon } x^{\varepsilon  + 1} \left( {x\left( {\frac{{s_{23} }}{q} - 1} \right) + 1} \right)^{3\varepsilon } 
\]
and the DE
\be
\begin{array}{ll}
 \frac{\partial }{{\partial x}}\left( {M_{001011011} G_{001011011} } \right) = -( - q)^{3\varepsilon } x^{\varepsilon  -2} \left( {x\left( {\frac{{s_{23} }}{q} - 1} \right) + 1} \right)^{3\varepsilon } \frac{1}{{s_{12} \varepsilon (2\varepsilon  - 1)\left( {q(x - 1) - s_{23} x} \right)}} \times  \\ 
 \left( {s_{12} x^2 \varepsilon \left( {x\varepsilon G_{001010012} \left( {q - 2s_{12} x + s_{12} } \right) - 2\left( {6\varepsilon ^2  - 5\varepsilon  + 1} \right)G_{001010011} } \right)} \right. \\ 
 \left. { + s_{12} x^2 \varepsilon \left( {6\varepsilon ^2  - 5\varepsilon  + 1} \right)G_{001001011}  + \left( {18\varepsilon ^3  - 27\varepsilon ^2  + 13\varepsilon  - 2} \right)G_{001010001} } \right) \\ 
 \end{array}
\ee
involves the 3 off-shell-legs triangles already calculated before, along with a 2-off-shell-legs triangle and a two-point MI.
The 2-off-shell-legs triangle satisfies of course its own DE expressed only in terns of two-point MI. The solution is again straightforward and given by 
\be
\begin{array}{l}
 M_{{\rm{001001011}}} G_{{\rm{001001011}}}  = A_3 \left( {\frac{{( - q)^{ - \varepsilon } \left( { - s_{12}  - s_{23} } \right)^\varepsilon  x^\varepsilon  }}{{\varepsilon ^2 \left( {2\varepsilon  - 1} \right)}}} \right. +  \\ 
 \frac{1}{\varepsilon }\left( {G\left( {\frac{q}{{q - s_{23} }},x} \right) - G\left( {\frac{{s_{12}  + s_{23} }}{{s_{12} }},x} \right) - \frac{{3( - q)^{ - \varepsilon } \left( { - s_{12}  - s_{23} } \right)^\varepsilon  x^\varepsilon  }}{{2(2\varepsilon  - 1)}}} \right) +  \\ 
 2G\left( {0,\frac{q}{{s_{12} }},x} \right) - 2G\left( {0,\frac{q}{{q - s_{23} }},x} \right) - 2G\left( {\frac{q}{{q - s_{23} }},1,x} \right) - 2G\left( {\frac{q}{{q - s_{23} }},\frac{q}{{s_{12} }},x} \right) \\ 
  + G\left( {\frac{q}{{q - s_{23} }},\frac{q}{{q - s_{23} }},x} \right) + G\left( {\frac{q}{{q - s_{23} }},\frac{{s_{12}  + s_{23} }}{{s_{12} }},x} \right) + 2G\left( {\frac{{s_{12}  + s_{23} }}{{s_{12} }},\frac{q}{{s_{12} }},x} \right) \\ 
  - G\left( {\frac{{s_{12}  + s_{23} }}{{s_{12} }},\frac{q}{{q - s_{23} }},x} \right) + G\left( {\frac{{s_{12}  + s_{23} }}{{s_{12} }},x} \right)\left( {\log ( - q) - \log \left( { - s_{12}  - s_{23} } \right) - \log (x) - \frac{1}{2}} \right) \\ 
  + \left( { - \log ( - q) + \log \left( { - s_{12}  - s_{23} } \right) + \log (x) + \frac{1}{2}} \right)G\left( {\frac{q}{{q - s_{23} }},x} \right) + 2G\left( {\frac{{s_{12}  + s_{23} }}{{s_{12} }},1,x} \right) \\ 
 \left. { - G\left( {\frac{{s_{12}  + s_{23} }}{{s_{12} }},\frac{{s_{12}  + s_{23} }}{{s_{12} }},x} \right) + 2G(0,1,x) + ( - q)^{ - \varepsilon } \left( { - \left( { - s_{12}  - s_{23} } \right)^\varepsilon  } \right)x^\varepsilon  } \right) \\ 
 \end{array}
\ee
Having now all the ingredients we can solve the DE for the hard box, resulting to
{\tiny
\be
\begin{array}{l}
 M_{{\rm{001011011}}} G_{{\rm{001011011}}}  = \frac{{A_3 }}{{q - s_{23} }}\left( {\frac{2}{{\varepsilon ^2 }}G\left( {\frac{q}{{q - s_{23} }},x} \right)} \right. +  \\ 
 \frac{1}{\varepsilon }\left( {2G\left( {0,\frac{q}{{q - s_{23} }},x} \right) + 8G\left( {\frac{q}{{q - s_{23} }},\frac{q}{{q - s_{23} }},x} \right) + \left( {4\log ( - q) - 2\log \left( { - s_{12} } \right) - 2\log (x) - 5} \right)G\left( {\frac{q}{{q - s_{23} }},x} \right)} \right) +  \\ 
 10G\left( {0,0,\frac{q}{{q - s_{23} }},x} \right) + 4G\left( {0,\frac{q}{{q - s_{23} }},1,x} \right) + 4G\left( {0,\frac{q}{{q - s_{23} }},\frac{q}{{s_{12} }},x} \right) + 4G\left( {0,\frac{q}{{q - s_{23} }},\frac{q}{{q - s_{23} }},x} \right) +  \\ 
 2G\left( {\frac{q}{{q - s_{23} }},0,1,x} \right) + 2G\left( {\frac{q}{{q - s_{23} }},0,\frac{q}{{s_{12} }},x} \right) - 2G\left( {\frac{q}{{q - s_{23} }},0,\frac{q}{{q - s_{23} }},x} \right) - 2G\left( {\frac{q}{{q - s_{23} }},\frac{q}{{q - s_{23} }},1,x} \right) -  \\ 
 2G\left( {\frac{q}{{q - s_{23} }},\frac{q}{{q - s_{23} }},\frac{q}{{s_{12} }},x} \right) + 28G\left( {\frac{q}{{q - s_{23} }},\frac{q}{{q - s_{23} }},\frac{q}{{q - s_{23} }},x} \right) + 2G\left( {\frac{q}{{q - s_{23} }},\frac{{s_{12}  + s_{23} }}{{s_{12} }},1,x} \right) +  \\ 
 2G\left( {\frac{q}{{q - s_{23} }},\frac{{s_{12}  + s_{23} }}{{s_{12} }},\frac{q}{{s_{12} }},x} \right) - 2G\left( {\frac{q}{{q - s_{23} }},\frac{{s_{12}  + s_{23} }}{{s_{12} }},\frac{q}{{q - s_{23} }},x} \right) +  \\ 
 \left( {\log (x)\left( { - 8\log ( - q) + 6\log \left( { - s_{12} } \right) + 5} \right) - 2\log ( - q)\left( {3\log \left( { - s_{12} } \right) + 5} \right) + 5\log ^2 ( - q) + 2\log ^2 \left( { - s_{12} } \right) + 5\log \left( { - s_{12} } \right) - 5\log ^2 (x) - 1} \right)G\left( {\frac{q}{{q - s_{23} }},x} \right) \\ 
  + \left( {8\log ( - q) - 6\log \left( { - s_{12} } \right) - 5} \right)G\left( {0,\frac{q}{{q - s_{23} }},x} \right) + \left( {2\left( {\log ( - q) - \log \left( { - s_{12} } \right)} \right) - 4\log (x)} \right)G\left( {\frac{q}{{q - s_{23} }},1,x} \right) +  \\ 
 \left( {2\left( {\log ( - q) - \log \left( { - s_{12} } \right)} \right) - 4\log (x)} \right)G\left( {\frac{q}{{q - s_{23} }},\frac{q}{{s_{12} }},x} \right) + \left( {14\log ( - q) - 6\log \left( { - s_{12} } \right) - 4\log (x) - 20} \right)G\left( {\frac{q}{{q - s_{23} }},\frac{q}{{q - s_{23} }},x} \right) +  \\ 
\left. 10\log (x)G\left( {\frac{q}{{q - s_{23} }},0,x} \right) \right)
 \end{array}
\ee
}

\begin{figure}
\begin{center}
\includegraphics[width=8cm]{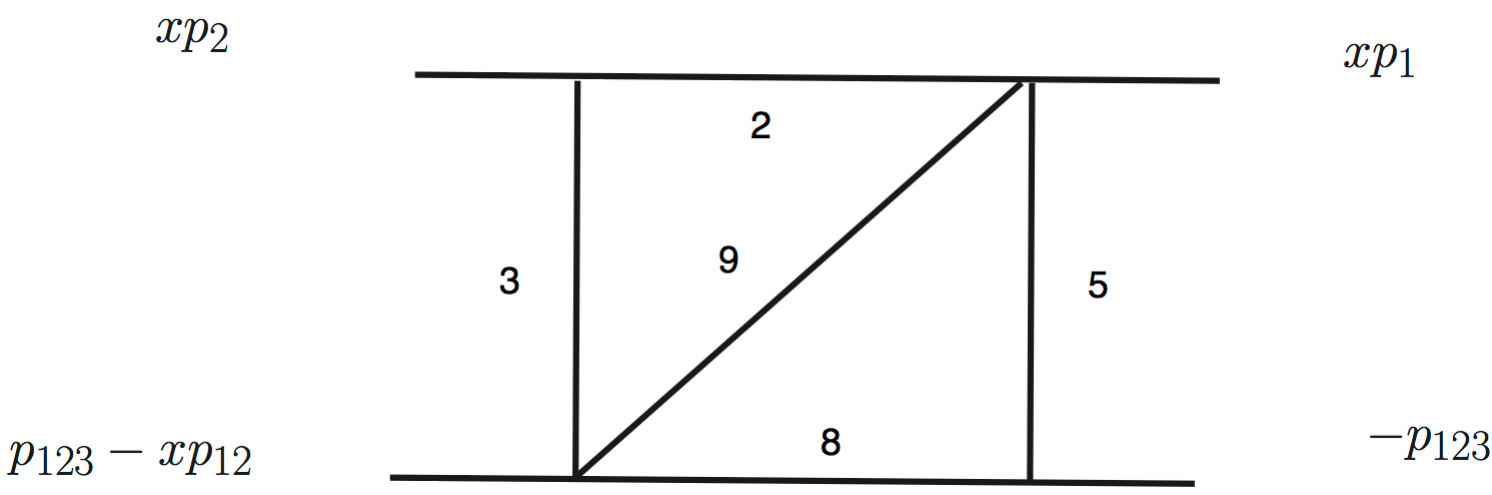}
%\[ \epsfig{figure=boxd.eps,,width=8cm} \]
\caption{Two-loop box MI with two off-shell legs: the diagonal hard one, $G_{011010011}$. See in the text for more details. Labels as in Fig.~\ref{fig:boxd}.} 
\label{fig:boxd}
\end{center}
\end{figure}

Finally we consider the diagonal box with two adjacent off-shell legs, given in Fig.~\ref{fig:boxd}.
There are two MI associated, namely $G_{011010011}$ and $G_{011010021}$.
The DE are given by  

\[
\begin{array}{ll}
 \frac{\partial }{{\partial x}}G_{011010011}  &=  - \frac{{\left( {9d^3  - 81d^2  + 242d - 240} \right)}}{{(d - 4)^2 (x - 1)x^2 \left( {s_{12} ( - x) + s_{12}  + s_{23} } \right)\left( {q - s_{12} x} \right)}}G_{001000011}  \\ 
  &- \frac{{\left( {9d^3  - 81d^2  + 242d - 240} \right)}}{{(d - 4)^2 x^2 \left( {s_{12} ( - x) + s_{12}  + s_{23} } \right)\left( {q( - x) + q + s_{23} x} \right)}}G_{010000011}  + \frac{{(2d - 9)}}{x}G_{011010011}  - \frac{q}{x}G_{011010021}  
 \end{array}
\]
%and

{\footnotesize
\[
\begin{array}{ll}
 \frac{\partial }{{\partial x}}G_{011010021}  &= \frac{{\left( {9d^3  - 81d^2  + 242d - 240} \right)\left( {3q^2 (x - 1) - q\left( {s_{12} \left( {x^2  + x - 2} \right) + s_{23} (3x - 5)} \right) + s_{12} s_{23} (x - 3)x} \right)}}{{2(d - 4)q(x - 1)^2 x\left( {s_{12} ( - x) + s_{12}  + s_{23} } \right)\left( {q - s_{12} x} \right)^2 \left( {q(x - 1) - s_{23} x} \right)}}G_{001000011}  \\ 
  &+ \frac{{\left( {9d^3  - 81d^2  + 242d - 240} \right)\left( {q\left( {s_{12} (x - 1)^2  + s_{23} } \right) - s_{12} s_{23} x^2 } \right)}}{{2(d - 4)qs_{12} (x - 1)x^3 \left( {s_{12} ( - x) + s_{12}  + s_{23} } \right)\left( {q - s_{12} x} \right)\left( {q(x - 1) - s_{23} x} \right)}}G_{001010001}  
\\ 
 & + \frac{{\left( {3d^2  - 19d + 30} \right)\left( {3q^2 (x - 1) + q\left( {s_{12} \left( { - 2x^2  + x + 1} \right) + s_{23} (4 - 3x)} \right) + s_{12} s_{23} x(2x - 3)} \right)}}{{2q(x - 1)x\left( {s_{12} ( - x) + s_{12}  + s_{23} } \right)\left( {s_{12} x - q} \right)\left( {q( - x) + q + s_{23} x} \right)}}G_{001010011}  
\\ 
  &- \frac{{(d - 4)\left( {q^2 \left( {2s_{12} (x - 1)^2  - s_{23} } \right) - qs_{12} \left( {s_{23} \left( {2x^2  - 5x + 1} \right) + s_{12} (x + 1)(x - 1)^2 } \right) + s_{12} ^2 s_{23} (x - 2)x^2 } \right)}}{{q(x - 1)\left( {s_{12} ( - x) + s_{12}  + s_{23} } \right)\left( {s_{12} x - q} \right)\left( {q( - x) + q + s_{23} x} \right)}}G_{001010012}  \\ 
  &- \frac{{3\left( {9d^3  - 81d^2  + 242d - 240} \right)\left( {q - s_{23} } \right)}}{{2(d - 4)qx\left( {s_{12} ( - x) + s_{12}  + s_{23} } \right)\left( {q(x - 1) - s_{23} x} \right)^2 }}G_{010000011}  \\ 
  &- \frac{{3\left( {3d^2  - 19d + 30} \right)\left( {q - s_{23} } \right)}}{{2qx\left( {s_{12} ( - x) + s_{12}  + s_{23} } \right)\left( {q(x - 1) - s_{23} x} \right)}}G_{010010011}  \\ 
  &+ \frac{{3(d - 4)^2 \left( {s_{12}  + s_{23} } \right)}}{{2x\left( {s_{12} ( - x) + s_{12}  + s_{23} } \right)\left( {q( - x) + q + s_{23} x} \right)}}G_{011010011}  \\ 
  &+ \frac{{\left( {q\left( {s_{12} (x - 1)(4dx + d - 20x - 2) - s_{23} (3dx + d - 14x - 2)} \right) + s_{23} x\left( {s_{12} ( - 4dx + 3d + 20x - 14) + (3d - 14)s_{23} } \right)} \right)}}{{2x\left( {s_{12} ( - x) + s_{12}  + s_{23} } \right)\left( {q( - x) + q + s_{23} x} \right)}}G_{011010021}  \\ 
 \end{array}
\]
}
As it is easily seen it involves already obtained results, such as the off-shell  and on-shell triangles.
The integrating factors are given by 
\[
M_{011010011}  = x^{4\varepsilon  + 1} 
\]
%and
{\footnotesize
\[
M_{011010021}  = ( - q)^{4\varepsilon } \left( {s_{12} x - s_{12}  - s_{23} } \right)\left( { - s_{12}  - s_{23} } \right)^\varepsilon  x^{1 - \varepsilon } \left( {1 - \frac{{s_{12} x}}{{s_{12}  + s_{23} }}} \right)^\varepsilon  \left( {x\left( {\frac{{s_{23} }}{q} - 1} \right) + 1} \right)^{4\varepsilon } 
\]
}
The result is given by ($u_1=(s_{12}+s_{23})/s_{12}$, $u_2=q/(q-s_{23})$, $u_3=q/s_{12}$),
\be
M_{011010011} G_{011010011}  = \frac{{2A_3 }}{{s_{12}  + s_{23} }}\sum\limits_{i =  - 1} {f_i \varepsilon ^i } 
\ee
{\tiny
\begin{dgroup*}
\begin{dmath*}
 f_{ - 1}  = \left( {\log ( - q) - \log \left( { - s_{12} } \right)} \right)G\left( {0,u_3 ,x} \right) + \left( {\log ( - q) - \log \left( { - s_{12} } \right)} \right)G\left( {u_1 ,u_2 ,x} \right) 
  + \left( {\log \left( { - s_{12} } \right) - \log ( - q)} \right)G\left( {0,u_2 ,x} \right) + \left( {\log \left( { - s_{12} } \right) - \log ( - q)} \right)G\left( {u_1 ,1,x} \right) 
  + \left( {\log \left( { - s_{12} } \right) - \log ( - q)} \right)G\left( {u_1 ,u_3 ,x} \right) + G(0,1,x)\left( {\log ( - q) - \log \left( { - s_{12} } \right)} \right) 
  + G\left( {0,0,u_2 ,x} \right) - G\left( {0,0,u_3 ,x} \right) + G\left( {0,u_1 ,1,x} \right) - G\left( {0,u_1 ,u_2 ,x} \right) 
  + G\left( {0,u_1 ,u_3 ,x} \right) 
  + 2G\left( {0,u_2 ,0,x} \right) - G\left( {0,u_2 ,1,x} \right) + G\left( {0,u_2 ,u_2 ,x} \right) 
  - G\left( {0,u_2 ,u_3 ,x} \right) 
%\end{dmath*}
%\begin{dmath*}  
%\left.
  - 2G\left( {0,u_3 ,0,x} \right) + G\left( {u_1 ,0,1,x} \right) - G\left( {u_1 ,0,u_2 ,x} \right) + G\left( {u_1 ,0,u_3 ,x} \right) + 2G\left( {u_1 ,1,0,x} \right) 
  - G\left( {u_1 ,u_1 ,1,x} \right) + G\left( {u_1 ,u_1 ,u_2 ,x} \right) - G\left( {u_1 ,u_1 ,u_3 ,x} \right) - 2G\left( {u_1 ,u_2 ,0,x} \right) 
+ G\left( {u_1 ,u_2 ,1,x} \right) 
  - G\left( {u_1 ,u_2 ,u_2 ,x} \right) + G\left( {u_1 ,u_2 ,u_3 ,x} \right) + 2G\left( {u_1 ,u_3 ,0,x} \right) - G(0,0,1,x) - 2G(0,1,0,x) 
%\right.
\end{dmath*}
%\begin{comment}
\begin{dmath*}
  f_0 = -9 G(0,0,0,1,x)+9 G\left(0,0,0,u_2,x\right)-9 G\left(0,0,0,u_3,x\right)-14 G(0,0,1,0,x)+2 G(0,0,1,1,x)+2 G\left(0,0,1,u_3,x\right)+5
   G\left(0,0,u_1,1,x\right)-5 G\left(0,0,u_1,u_2,x\right)+5 G\left(0,0,u_1,u_3,x\right)+14 G\left(0,0,u_2,0,x\right)-5
   G\left(0,0,u_2,1,x\right)+3 G\left(0,0,u_2,u_2,x\right)-5 G\left(0,0,u_2,u_3,x\right)-14 G\left(0,0,u_3,0,x\right)
%\end{dmath*}
%\begin{dmath*}
   +2
   G\left(0,0,u_3,1,x\right)+2 G\left(0,0,u_3,u_3,x\right)-8 G(0,1,0,0,x)+G(0,1,0,1,x)+G\left(0,1,0,u_3,x\right)+4 G(0,1,1,0,x)-2
   G\left(0,1,u_3,0,x\right)+5 G\left(0,u_1,0,1,x\right)-5 G\left(0,u_1,0,u_2,x\right)+5 G\left(0,u_1,0,u_3,x\right)+6
   G\left(0,u_1,1,0,x\right)-2 G\left(0,u_1,1,1,x\right)-2
   G\left(0,u_1,1,u_3,x\right)-G\left(0,u_1,u_1,1,x\right)+G\left(0,u_1,u_1,u_2,x\right)-G\left(0,u_1,u_1,u_3,x\right)-6
   G\left(0,u_1,u_2,0,x\right)+G\left(0,u_1,u_2,1,x\right)+G\left(0,u_1,u_2,u_2,x\right)+G\left(0,u_1,u_2,u_3,x\right)
   +6 G\left(0,u_1,u_3,0,x\right)-2 G\left(0,u_1,u_3,1,x\right)-2 G\left(0,u_1,u_3,u_3,x\right)+8 G\left(0,u_2,0,0,x\right)-4
   G\left(0,u_2,0,1,x\right)+3 G\left(0,u_2,0,u_2,x\right)-4 G\left(0,u_2,0,u_3,x\right)+2 G\left(0,u_2,1,1,x\right)
   +2
   G\left(0,u_2,1,u_3,x\right)-4 G\left(0,u_2,u_1,1,x\right)+4 G\left(0,u_2,u_1,u_2,x\right)-4 G\left(0,u_2,u_1,u_3,x\right)
   -4
   G\left(0,u_2,u_2,0,x\right)+4 G\left(0,u_2,u_2,1,x\right)-6 G\left(0,u_2,u_2,u_2,x\right)+4 G\left(0,u_2,u_2,u_3,x\right)+2
   G\left(0,u_2,u_3,1,x\right)+2 G\left(0,u_2,u_3,u_3,x\right)-8
   G\left(0,u_3,0,0,x\right)+G\left(0,u_3,0,1,x\right)+G\left(0,u_3,0,u_3,x\right)-2 G\left(0,u_3,1,0,x\right)+4
   G\left(0,u_3,u_3,0,x\right)+9 G\left(u_1,0,0,1,x\right)-9 G\left(u_1,0,0,u_2,x\right)+9 G\left(u_1,0,0,u_3,x\right)+14
   G\left(u_1,0,1,0,x\right)-2 G\left(u_1,0,1,1,x\right)-2 G\left(u_1,0,1,u_3,x\right)-5 G\left(u_1,0,u_1,1,x\right)+5
   G\left(u_1,0,u_1,u_2,x\right)-5 G\left(u_1,0,u_1,u_3,x\right)-14 G\left(u_1,0,u_2,0,x\right)
%\end{dmath*}
%\begin{dmath*}
+5 G\left(u_1,0,u_2,1,x\right)-3
   G\left(u_1,0,u_2,u_2,x\right)+5 G\left(u_1,0,u_2,u_3,x\right)+14 G\left(u_1,0,u_3,0,x\right)-2 G\left(u_1,0,u_3,1,x\right)-2
   G\left(u_1,0,u_3,u_3,x\right)+8 G\left(u_1,1,0,0,x\right)-G\left(u_1,1,0,1,x\right)-G\left(u_1,1,0,u_3,x\right)-4
   G\left(u_1,1,1,0,x\right)+2 G\left(u_1,1,u_3,0,x\right)-5 G\left(u_1,u_1,0,1,x\right)+5 G\left(u_1,u_1,0,u_2,x\right)-5
   G\left(u_1,u_1,0,u_3,x\right)-6 G\left(u_1,u_1,1,0,x\right)+2 G\left(u_1,u_1,1,1,x\right)+2
   G\left(u_1,u_1,1,u_3,x\right)+G\left(u_1,u_1,u_1,1,x\right)-G\left(u_1,u_1,u_1,u_2,x\right)+G\left(u_1,u_1,u_1,u_3,x\right)
+6
   G\left(u_1,u_1,u_2,0,x\right)-G\left(u_1,u_1,u_2,1,x\right)-G\left(u_1,u_1,u_2,u_2,x\right)-G\left(u_1,u_1,u_2,u_3,x\right)-6
   G\left(u_1,u_1,u_3,0,x\right)+2 G\left(u_1,u_1,u_3,1,x\right)
      +2 G\left(u_1,u_1,u_3,u_3,x\right)-8 G\left(u_1,u_2,0,0,x\right)+4
   G\left(u_1,u_2,0,1,x\right)-3 G\left(u_1,u_2,0,u_2,x\right)+4 G\left(u_1,u_2,0,u_3,x\right)-2 G\left(u_1,u_2,1,1,x\right)
   -2
   G\left(u_1,u_2,1,u_3,x\right)+4 G\left(u_1,u_2,u_1,1,x\right)-4 G\left(u_1,u_2,u_1,u_2,x\right)+4 G\left(u_1,u_2,u_1,u_3,x\right)+4
   G\left(u_1,u_2,u_2,0,x\right)-4 G\left(u_1,u_2,u_2,1,x\right)+6 G\left(u_1,u_2,u_2,u_2,x\right)-4 G\left(u_1,u_2,u_2,u_3,x\right)-2
   G\left(u_1,u_2,u_3,1,x\right)-2 G\left(u_1,u_2,u_3,u_3,x\right)
   +8
   G\left(u_1,u_3,0,0,x\right)-G\left(u_1,u_3,0,1,x\right)-G\left(u_1,u_3,0,u_3,x\right)+2 G\left(u_1,u_3,1,0,x\right)-4
   G\left(u_1,u_3,u_3,0,x\right)
   +G\left(0,u_2,0,x\right) (-4 \log (-q)-9)+G\left(u_1,1,0,x\right) (-4 \log
   (-q)-9)+G\left(u_1,u_3,0,x\right) (-4 \log (-q)-9)+G(0,1,0,x) (4 \log (-q)+9)+G\left(0,u_3,0,x\right) (4 \log
   (-q)+9)+G\left(u_1,u_2,0,x\right) (4 \log (-q)+9)
+G(0,0,1,x) \left(7 \log (-q)-5 \log
   \left(-s_{12}\right)+\frac{9}{2}\right)+G\left(0,0,u_3,x\right) \left(7 \log (-q)-5 \log
   \left(-s_{12}\right)+\frac{9}{2}\right)+G\left(u_1,0,u_2,x\right) \left(7 \log (-q)-5 \log
   \left(-s_{12}\right)+\frac{9}{2}\right)+G\left(0,u_2,u_2,x\right) \left(2 \log (-q)-4 \log
   \left(-s_{12}\right)-\frac{9}{2}\right)+G\left(u_1,u_2,1,x\right) \left(-2 \log
   \left(-s_{12}\right)-\frac{9}{2}\right)
     +G\left(u_1,u_2,u_3,x\right) \left(-2 \log
   \left(-s_{12}\right)-\frac{9}{2}\right)
+G\left(0,1,u_3,x\right) \left(\log (-q)-\log
   \left(-s_{12}\right)\right)+G\left(0,u_3,1,x\right) \left(\log (-q)-\log \left(-s_{12}\right)\right)+2 G\left(u_1,1,1,x\right)
   \left(\log (-q)-\log \left(-s_{12}\right)\right)+2 G\left(u_1,u_3,u_3,x\right) \left(\log (-q)-\log
   \left(-s_{12}\right)\right)+G\left(0,u_1,u_2,x\right) \left(3 \log (-q)-\log
   \left(-s_{12}\right)+\frac{9}{2}\right)+G\left(u_1,u_1,1,x\right) \left(3 \log (-q)-\log
   \left(-s_{12}\right)+\frac{9}{2}\right)+G\left(u_1,u_1,u_3,x\right) \left(3 \log (-q)-\log
   \left(-s_{12}\right)+\frac{9}{2}\right)+G\left(0,u_1,1,x\right) \left(-3 \log (-q)
   +\log
   \left(-s_{12}\right)
   -\frac{9}{2}\right)+G\left(0,u_1,u_3,x\right) \left(-3 \log (-q)+\log
   \left(-s_{12}\right)-\frac{9}{2}\right)+G\left(u_1,u_1,u_2,x\right) \left(-3 \log (-q)+\log \left(-s_{12}\right)-\frac{9}{2}\right)
%\end{dmath*}
%\begin{dmath*}
+2
   G(0,1,1,x) \left(\log \left(-s_{12}\right)-\log (-q)\right)+2 G\left(0,u_3,u_3,x\right) \left(\log \left(-s_{12}\right)-\log
   (-q)\right)+G\left(u_1,1,u_3,x\right) \left(\log \left(-s_{12}\right)-\log (-q)\right)+G\left(u_1,u_3,1,x\right) \left(\log
   \left(-s_{12}\right)-\log (-q)\right)+G\left(0,u_2,1,x\right) \left(2 \log
   \left(-s_{12}\right)+\frac{9}{2}\right)+G\left(0,u_2,u_3,x\right) \left(2 \log
   \left(-s_{12}\right)+\frac{9}{2}\right)+G\left(u_1,u_2,u_2,x\right) \left(-2 \log (-q)+4 \log
   \left(-s_{12}\right)+\frac{9}{2}\right)+G\left(0,0,u_2,x\right) \left(-7 \log (-q)
%\end{dmath*}
%\begin{dmath*}
%\left. 
   +5 \log
   \left(-s_{12}\right)
%\right.
   -\frac{9}{2}\right)+G\left(u_1,0,1,x\right) \left(-7 \log (-q)+5 \log
   \left(-s_{12}\right)-\frac{9}{2}\right)+G\left(u_1,0,u_3,x\right) \left(-7 \log (-q)+5 \log
   \left(-s_{12}\right)-\frac{9}{2}\right)+G(0,1,x) \left(-\log ^2(-q)-\frac{9 \log (-q)}{2}+\log ^2\left(-s_{12}\right)+\frac{9}{2} \log
   \left(-s_{12}\right)\right)+G\left(0,u_3,x\right) \left(-\log ^2(-q)-\frac{9 \log (-q)}{2}+\log ^2\left(-s_{12}\right)+\frac{9}{2}
   \log \left(-s_{12}\right)\right)+G\left(u_1,u_2,x\right) \left(-\log ^2(-q)-\frac{9 \log (-q)}{2}+\log
   ^2\left(-s_{12}\right)+\frac{9}{2} \log \left(-s_{12}\right)\right)+G\left(0,u_2,x\right) \left(\log ^2(-q)+\frac{9 \log
   (-q)}{2}-\frac{1}{2} \log \left(-s_{12}\right) \left(2 \log \left(-s_{12}\right)+9\right)\right)+G\left(u_1,1,x\right) \left(\log
   ^2(-q)+\frac{9 \log (-q)}{2}-\frac{1}{2} \log \left(-s_{12}\right) \left(2 \log
   \left(-s_{12}\right)+9\right)\right)+G\left(u_1,u_3,x\right) \left(\log ^2(-q)+\frac{9 \log (-q)}{2}-\frac{1}{2} \log
   \left(-s_{12}\right) \left(2 \log \left(-s_{12}\right)+9\right)\right)
\end{dmath*}
%\end{comment}
\end{dgroup*} 
}

All results have been numerically tested for euclidean kinematics\footnote{Tests have also been successfully performed for physical kinematics, but a detailed discussion will be given in a forthcoming publication.} 
against {\tt SecDec} package~\cite{Carter:2010hi,Borowka:2012yc}, always with at least 6 digits accuracy,
whereas our numerical evaluation of GPs was tested against the package provided in Ginac~\cite{Bauer:2000cp} developed in reference~\cite{Vollinga:2004sn}. 

\section{Discussion}
\label{section:Di}

The simplified DE approach presented in this paper may provide a new path in calculating the MI at two loops needed for physical applications. 
It allows for a straightforward derivation of DE in a parametrisation suitable to express directly the MI in terms of GPs, at least for MI with vanishing internal masses. 
We have also noticed that,  in almost all cases, the DE obtained 
can give the full answer without the need of an independent evaluation of the boundary conditions. As a proof-of-concept of the proposed method, we have derived several known and new results
at one- and two-loop level.

The knowledge of two--loop MI combined with the extension of the OPP method at two loops
and the use of IBPI 
will pave the road towards an automatization of NNLO calculations, within a full numerical framework, similarly to what  has already been achieved at NLO. 
MI for planar and non-planar double boxes and pentaboxes exhibit more complicated structure. 
Nevertheless, the experience accumulated so far, shows that a library of analytic expressions of all MI, at least for those with vanishing internal masses, needed for any two-loop amplitude is feasible. 
We intend to pursue this road in the near future.
Moreover, there are several issues that have to be understood better, among them:
\begin{itemize}
\item It is important to understand the relation of the parametrization introduced in this paper with the usual DE approach as well as to other approaches~\cite{Henn:2013pwa,Henn:2013woa,Henn:2013nsa,Argeri:2014qva}, especially
for systems of coupled  DE. 
It seems that the interplay with IPBI plays an important role. It is therefore very welcome
to exploit this issue in detail.
\item Although we found that in almost all cases no need of an independent calculation of the boundary conditions is needed, it will be safer and very welcome to have an independent method to calculate the $x=0$ limit\footnote{For 
asymptotic expansions of Feynman Integrals see ref.~\cite{Smirnov:2002pj} and references therein} of a given MI.
This is the subject of a current project.
\item DE introduced in this paper have the nice property, that their results are straightforwardly expressible in terms of GPs. Nevertheless, as the one-loop 5-point result suggests, expressions may be quite involved. One cannot exclude the possibility of further
simplifications by systematic use of GPs identities and symbol algebra. On the other hand a thorough analysis of numerical implementation is needed in order to assess the full potential of the approach for physical applications. 
In the same lines, the extension into the physical region kinematics should be studied, based on the $i\varepsilon$ prescription for the kinematical invariants involved and the analytic properties of the GPs.
\item In order to attack more cases of physical interest, MI involving internal masses should also be considered. A very preliminary study suggest that the parametrization should then be also associated to the internal masses.
We plan to further elaborate on that issue in the near future. 
\end{itemize}

%\noindent {\bf Acknowledgement:} 
\acknowledgments

I would like to acknowledge discussions with C.~Anastasiou, A.~Lazopoulos, C.~Duhr, G.~Heinrich, M.~Czakon, T.~Gehrmann, T.~Hahn, F.~Chavez, C.~Wever, D.~Tommasini. 
I would also like to thank CERN Theory Division and the Theoretical Physics Department of ETH, for their kind hospitality during the progress of this work.
This research has  been partially supported by the Research Funding Program "ARISTEIA", "Higher Order Calculations
and Tools for High Energy Colliders", HOCTools (co-financed by the European Union (European
Social Fund ESF) and Greek national funds through the Operational Program "Education
and Lifelong Learning" of the National Strategic Reference Framework (NSRF)).

\bibliographystyle{JHEP}   
\bibliography{Biblio}     

\end{document}